\documentclass[preprint,12pt]{elsarticle}

\usepackage{amssymb}
\usepackage{amsmath}
\usepackage{multirow}
\usepackage{float}
\usepackage{bm}
\usepackage{graphicx}
\usepackage{caption}
\usepackage{subcaption}
\usepackage{natbib}
\usepackage{pmat}
\usepackage{nomencl}
\usepackage{epstopdf}
\makenomenclature

\usepackage{lineno}

\journal{Energy}

\begin{document}

\begin{frontmatter}



\author[lneg]{Teresa Scholz}
\ead{teresa.scholz@lneg.pt}
\author[lneg]{Vitor V. Lopes\corref{cor1}}
\ead{vitor.lopes@lneg.pt}
\author[lneg]{Ana Estanqueiro}
\ead{ana.estanqueiro@lneg.pt}
\address[lneg]{LNEG, National Laboratory for Energy and Geology,
Estrada do Pa\c{c}o do Lumiar, 22, 1649-038, Lisboa, Portugal}

\cortext[cor1]{Corresponding author}

\title{A cyclic time-dependent Markov process to model daily patterns in wind turbine power production}


\begin{abstract}

Wind energy is becoming a top contributor to the renewable energy mix, which raises potential reliability issues for the grid due to the fluctuating nature of its source. To achieve adequate reserve commitment and to promote market participation, it is necessary to provide models that can capture daily patterns in wind power production. This paper presents a cyclic inhomogeneous Markov process, which is based on a three-dimensional state-space (wind power, speed and direction). Each time-dependent transition probability is expressed as a Bernstein polynomial. The model parameters are estimated by solving a constrained optimization problem: The objective function combines two maximum likelihood estimators, one to ensure that the Markov process long-term behavior reproduces the data accurately and another to capture daily fluctuations. A convex formulation for the overall optimization problem is presented and its applicability demonstrated through the analysis of a case-study. The proposed model is capable of 
reproducing the diurnal patterns of a three-year dataset collected from a wind turbine located in a mountainous region in Portugal. In addition, it is shown how to compute persistence statistics directly from the Markov process transition matrices. Based on the case-study, the power production persistence through the daily cycle is analysed and discussed.

\end{abstract}


\end{frontmatter}

\linenumbers

\newpage
\section{Introduction}
\label{Introduction}


The EC European Parliament objective to achieve 20\% of the consumed energy from the renewable energy sector by 2020 introduced a serious challenge to the planning and operating of power systems. Wind energy is becoming a top contributor to the renewable energy mix due to rather high capacities and generation costs that are becoming competitive with conventional energy sources \citep{Wen2009}. However, wind energy systems suffer from a major drawback, the fluctuating nature of their source, which affects the grid security, the power system operation and market economics. There are several tools to deal with these issues, such as the knowledge of wind power persistence and wind speed or power simulation. Persistence is related to stability properties and can provide useful information for bidding on the electricity market or to maintain reliability, e.g. by setting reserve capacity.


Wind power or speed simulation can be used to study the impact of wind generation on the power system. For this task, a sufficiently long time series of the power output from the wind plants should be used. However, real data records are commonly of short length and thus synthetic time series are generated by stochastic simulation techniques to model wind activity \citep{Papaefthymiou2008}.
\citet{Shamshad2005} used first and second-order Markov chain models for the generation of hourly wind speed time series. They found that a model with 12 wind speed states (1 m/s size) can capture the shape of the probability density function and preserve the properties of the observed time series. Additionally, they concluded that a second-order Markov chain produces better results. \citet{Nfaoui2004} compared the limiting behavior of their Markov chain model with the data histograms gotten from hourly averaged wind speed and showed that the statistical characteristics were faithfully reproduced. \citet{Sahin2001} reported the use of a first-order Markov chain approach to simulate the wind speed, where: a) both transitions between consecutive times and within state wind speeds are sampled using an uniform distribution; and, b) extreme states are sampled with an exponential distribution. They showed that statistical parameters were preserved to a significant extent; however, second-order Markov chain 
models could yield improved results.

Although wind power can be computed from synthetic wind speed time series, \citet{Papaefthymiou2008} show that a stochastic model using wind power leads to a reduced number of states and a lower Markov chain model order. They compared a Markov chain based method for the direct generation of wind power time series with the transformed generated wind speed. Both the autocorrelation and the probability density function of the simulated data showed a good fit. Thus, they concluded that it is better to generate wind power time series.
\citet{Chen2009} also modeled wind power by using different discrete Markov chain models: the basic Markov model; the Bayesian Markov model, which considers the transition matrix uncertainty; and, the birth-and-death Markov model, which only allows state transitions between immediately adjacent states. After comparing the wind power autocorrelation function, the authors find the Bayesian Markov model best.
\citet{Lopes2012} proposed a Markov chain model using states that combine information about wind speed, direction and power. From the transition matrix, they compute statistics, such as the stationary power distribution and persistence of power production, which show a close agreement with their empirical analogues. The model was then used for the two-dimensional stochastic modeling of wind dynamics by \citet{Raischel2012}. They aim at studying the interactions between wind velocity, turbine aerodynamics and controller action using a system of coupled stochastic equations describing the co-evolution of wind power and speed. They showed that both the deterministic and stochastic terms of the equations can be extracted directly from the Markov chain model.


The knowledge of wind power production persistence provides useful information to run a wind park and to bid on the electricity market, since it provides information about the expected power steadiness. It can be seen as the average time that a system remains in a given state or a subset of states. Existent literature focuses mainly on wind speed persistence, which is used for assessing the wind power potential of a region. Persistence can be determined directly from the data \citep{Pryor2002, Poje1992}; however, the presence of missing data leads to an underestimate of actual persistence. Alternative methods are based on wind speed duration curves \citep{Masseran2012, Koccak2002}, the autocorrelation function or conditional probabilities. \citet{Koccak2008} and \citet{Cancino-Solorzano2010} compare these techniques, and both conclude that wind speed duration curve yields the best results, i.e. results that follow the geographical and climatic conditions of the analyzed sites. Moreover, \citet{Cancino-Solorzano2010} analyze the concept of ``useful persistence'', which is the time schedule series where the wind speed is between the turbine cut-in and cut-out speed. The results gotten from this analysis coincide with the persistence classification obtained using the speed duration curves. In addition, \citet{Koccak2009} suggests a detrended fluctuation analysis to detect long-term correlations and analyze the persistence properties of wind speed records. \citet{Sigl1978, Corotis1978} and \citet{Poje1992} proposed an approach based on the use of a power law or exponential probability distributions for the persistence of wind speed above and below a reference value. A Markov chain based method to derive the distribution of persistence is introduced by \citet{Anastasiou1996}, who show its capability on wind speed data.


Most methods in literature of wind speed and power synthesis fail to represent diurnal patterns in the artificial data. However, these are relevant for energy system modeling and design, since their knowledge allows to plan and schedule better. For instance, a power production behavior that best matches demand needs smaller reserve capacity.
Recently, \citet{Suomalainen2012, Suomalainen2013} introduced a method for synthetic generation of wind speed scenarios that include daily wind patterns by sampling a probability distribution matrix based on five selected daily patterns and the mean speed of each day. \citet{Carapellucci2013} adopt a physical-statistical approach to synthesize wind speed data and evaluate the influence of the diurnal wind speed profile on the cross-correlation between produced energy and electrical loads. The parameters of their model, such as diurnal pattern strength or peak hour of wind speed are determined through a multi-objective optimization, carried out using a genetic algorithm.


This paper introduces a cyclic time-variant Markov model of wind power, speed and direction designed to consider the daily patterns observed in the data. The model can be used to synthesize data for the three variables and is capable of reproducing the daily patterns. Moreover, it allows to compute persistence statistics depending on the time of the day. The paper is organized as follows: Section~\ref{Theory} introduces the proposed model as an extension of the ``regular'' Markov chain model, which is then used for comparison. Furthermore it is shown, how to compute the time-of-the-day dependent persistence statistics directly from the Markov model transition matrices. In section~\ref{ParameterEstimation} the constrained convex optimization problem to get the model parameters is introduced and explained. It is applied to the analysis of a case-study based on real dataset, section~\ref{Example}. Since the model describes the joint statistics for wind power, speed and direction, Section~\ref{Simulation} 
explains how to create synthetic time-series for these variables. Section~{\ref{Results}} compares the synthesized data of both the time-variant and the time-invariant versions of the model. Moreover, it is shown how the persistence of power production varies through the daily cycle.

\section*{}
\begin{thenomenclature} 

 \nomgroup{A}

  \item [{$\alpha_0$}]\begingroup Initial state distribution at time step $t = 0$\nomeqref {1}
        \nompageref{6}
  \item [{$\beta_{\mu}^{i,j}$}]\begingroup Coefficients of the Bernstein polynomial modeling the transition probability $p_{i,j}(t)$\nomeqref {16}
        \nompageref{11}
  \item [{$\mathbf{1}_\mathcal{A}$}]\begingroup unit column vector of the same size as subset $\mathcal{A}$\nomeqref {4}
        \nompageref{7}
  \item [{$\mathbf{P}$}]\begingroup $P_0 \cdot \dotso \cdot P_{T-1}$\nomeqref {1}
        \nompageref{6}
  \item [{$\mathcal{A}$}]\begingroup Subset of the state space, containing the states of interest for persistence\nomeqref {2}
        \nompageref{7}
  \item [{$\mathcal{S}$}]\begingroup Set of observed state transitions\nomeqref {12}
        \nompageref{10}
  \item [{$\mathcal{S}_z$}]\begingroup Set of transitions observed in the data together with the scaled time of the day $z$ at which they are observed\nomeqref {16}
        \nompageref{12}
  \item [{$\omega$}]\begingroup Weight of the extra transitions added to the objective function\nomeqref {35}
        \nompageref{19}
  \item [{$\pi$}]\begingroup Stationary distribution of a time-invariant Markov chain\nomeqref {1}
        \nompageref{6}
  \item [{$\pi^{\ast}$}]\begingroup $\lim_{t \to \infty} \mathbf{P}^t$\nomeqref {1}
        \nompageref{6}
  \item [{$\pi_r$}]\begingroup Stationary distribution at time $r$ of a time-variant cyclic Markov process\nomeqref {1}
        \nompageref{6}
  \item [{$\pi_r(j)$}]\begingroup Stationary probability, of state $j$ at time of the day $r$\nomeqref {8}
        \nompageref{8}
  \item [{$\pi_{r}(\mathcal{A})$}]\begingroup Vector whose elements are the stationary probabilities of the states in the set $\mathcal{A}$ at time of the day $r$\nomeqref {9}
        \nompageref{9}
  \item [{$\tau$}]\begingroup Persistence\nomeqref {2}\nompageref{6}
  \item [{$\tau_r$}]\begingroup Time-dependent persistence in a cyclic Markov process\nomeqref {4}
        \nompageref{7}
  \item [{$b_{\mu, k}(z)$}]\begingroup $\mu$-th Bernstein basis polynomial of order $k$\nomeqref {14}
        \nompageref{11}
  \item [{$E[$ $]$}]\begingroup Expected value operator\nomeqref {4}
        \nompageref{7}
  \item [{$P_t$}]\begingroup $t$-th step transition matrix of a Markov process\nomeqref {1}
        \nompageref{5}
  \item [{$p_{i,j}(t)$}]\begingroup $t$-th step transition probability of a Markov process\nomeqref {1}
        \nompageref{5}
  \item [{$p_{i,j}^{avg}$}]\begingroup Daily average probability of transition from state $s_i$ to $s_j$\nomeqref {16}
        \nompageref{12}
  \item [{$r_t$}]\begingroup Remainder of time step $t$ modulo $T$\nomeqref {1}
        \nompageref{5}
  \item [{$S$}]\begingroup Markov process state space\nomeqref {0}
        \nompageref{5}
  \item [{$s_i$}]\begingroup $i$-th state of a Markov process\nomeqref {0}
        \nompageref{5}
  \item [{$T$}]\begingroup Period of a cyclic Markov process\nomeqref {1}
        \nompageref{5}
  \item [{$t$}]\begingroup Time step of a Markov process\nomeqref {0}
        \nompageref{5}
  \item [{$X_t$}]\begingroup Markov process\nomeqref {0}\nompageref{5}
  \item [{$z$}]\begingroup Scaled time of the day\nomeqref {16}
        \nompageref{11}
  \item [{${\pi}_{\mathcal{A}}$}]\begingroup Stationary probability distribution of the states in subset $\mathcal{A}$\nomeqref {4}
        \nompageref{7}
  \item [{r}]\begingroup time of the day\nomeqref {1}\nompageref{5}

\end{thenomenclature}

\newpage
\section{Time-inhomogeneous Markov model}
\label{Theory}

\subsection{Definition}

A discrete finite Markov process $\{X_t \in S, t \geq 0\}$\nomenclature{$X_t$}{Markov process} is a stochastic process on a discrete finite state space $S = \{s_0, s_1, ..., s_n\}$,~$n \in \mathbb{N}$\nomenclature{$S$}{Markov process state space}\nomenclature{$t$}{Time step of a Markov process}\nomenclature{$s_i$}{$i$-th state of a Markov process}, whose future evolution depends only on its current state \citep{Kemeny1976}. This Markov property is expressed mathematically by

\begin{small}
\begin{equation*}
Pr\{X_{t+1} = s_j \mid X_{t} = s_i \land X_l \in S \quad \forall l = 0, ..., t-1\} = Pr\{X_{t+1} = s_j\mid X_{t} = s_i\}.
\end{equation*}
\end{small}

$Pr\{X_{t+1} = s_j\mid X_{t} = s_i\}$\nomenclature{$p_{i,j}(t)$}{$t$-th step transition probability of a Markov process} describes the probability of the Markov process moving to state $s_j$ at time step $t+1$ given that it is in state $s_i$ and is called the $t$-th step transition probability, denoted as $p_{i,j}(t)$. Thus, for each time step $t$ the Markov process has an associated transition probability matrix $P_t$\nomenclature{$P_t$}{$t$-th step transition matrix of a Markov process}, a $n$ by $n$ matrix with entries $[P_t]_{i,j} = p_{i,j}(t)$ for all $i,j \in \lbrace 0,\ldots, n\rbrace$. Each $P_t$ satisfies the following properties: $p_{i,j}(t) \geq 0$ and $\sum_{j}{p_{i,j}(t)} = 1$ $\forall i,j \in \lbrace 0,\ldots, n\rbrace$,~$\forall t$. A Markov process is called cyclic with period $T \in \mathbb{N}$\nomenclature{$T$}{Period of a cyclic Markov process}, if $T$ is the smallest number, such that $p_{i,j}(mT + r) = p_{i,j}(r)$ for all $m$ in $\mathbb{N}$,~$0 \leq r < T$ \cite{Platis1998}. 
Thus, a cyclic Markov process is described by $T$ transition matrices $P_r$,~$r = 0,...,T-1$\nomenclature{r}{time of the day}. The remainder of time step $t$ modulo $T$ will be denoted as $r_t$\nomenclature{$r_t$}{Remainder of time step $t$ modulo $T$} and thus $r_t$ = $r_{t+mT}$~$\forall t, m \in \mathbb{N}$.\\
If the transition probabilities are time-independent, i.e. $p_{i,j}(t) = p_{i,j}$, the process is called a (time-homogeneous) Markov chain and its probability matrix $P \in \mathbb{R}^{n+1 \times n+1}$ is given by $[P]_{i,j} = p_{i,j}$. By analogy to the time-dependent case it holds that $p_{i,j} \geq 0$ and $\sum_{j}{p_{i,j}} = 1$ $\forall i,j \in \lbrace 0,\ldots, n\rbrace$.\\ 

\subsection{Communication classes and irreducibility}
\subsubsection{Time-invariant Markov chain}
The probability of reaching a state $s_j$ from a state $s_i$ in $l$ time steps is given by $P^l(i,j)$, i.e. the $l$-th power of the transition matrix $P$. If a state $s_j$ can be reached from a state $s_i$ in a finite number of time steps and vice versa, i.e. $\exists l \in \mathbb{N}$~$P^l(i,j) \geq 0 \land P^l(j,i) \geq 0$, the states $s_i$ and $s_j$ communicate. All states that communicate with each other are said to be in the same communication class. If all states of a state space are in the same communication class, i.e. if it is possible to reach every state from any other state in a finite number of time steps, the corresponding transition matrix $P$ is called irreducible.
\subsubsection{Cyclic time-variant Markov process}
A cyclic Markov process with period $T$ is described by $T$ transition matrices $P_r$, one for each time of the day $r = 0, ..., T-1$. The probability of the process reaching state $s_j$ from state $s_i$ in $l$ time steps at time $t = 0$ is given as $(P_0 \cdot \dotso \cdot P_{T-1})^m \cdot P_0 \cdot \dotso \cdot P_{r_l}(i,j)$ with $l = mT + r_l$. For an arbitrary time-step $t$, the formula must be multiplied from the left with the term $P_{r_t} \cdot \dotso \cdot P_{T-1}$. Thus, the Markov process is irreducible, if the matrix $\mathbf{P} = P_0 \cdot \dotso \cdot P_{T-1}$\nomenclature{$\mathbf{P}$}{$P_0 \cdot \dotso \cdot P_{T-1}$} is irreducible, i.e. if $\exists l \in \mathbb{N}$~$\mathbf{P}^l(i,j) \geq 0$~$\forall i,j$.

\subsection{The stationary distribution}

If a Markov chain is irreducible and aperiodic then the long-term statistics of a Markov chain are described by the stationary probability distribution: $\pi = \lim_{t \to \infty} \alpha_0 P^t$\nomenclature{$\pi$}{Stationary distribution of a time-invariant Markov chain}. The distribution is independent of the initial distribution $\alpha_0$\nomenclature{$\alpha_0$}{Initial state distribution at time step $t = 0$} and satisfies the balance equation $\pi = \pi P$. By the Perron-Frobenius theorem it can be computed as the normalized eigenvector corresponding to the eigenvalue $1$ of the transition matrix \citep{Pillai2005}.\\
In the case of the cyclic time-inhomogeneous Markov process there is also a stationary distribution $\pi_r$\nomenclature{$\pi_r$}{Stationary distribution at time $r$ of a time-variant cyclic Markov process}, for all $r < T$. It can be interpreted as the limiting distribution of the Markov process considering only the datapoints sampled at time of the day $r$. 
If the matrix $\mathbf{P}$ is irreducible, i.e. if $\exists \pi^{\ast}$\nomenclature{$\pi^{\ast}$}{$\lim_{t \to \infty} \mathbf{P}^t$}, such that $\pi^{\ast} = \lim_{t \to \infty} \alpha_0 \cdot \mathbf{P}^t$ and the process is aperiodic, the stationary distribution $\pi_r$ exists and is given by $\pi^{\ast}\cdot P_0 \cdot \dotso \cdot P_{r-1}$, since it satisfies the balance equation~\ref{balanceeq}:
\begin{align}
\pi_r &= \pi^{\ast} \cdot P_0 \cdot \dotso \cdot P_{r-1} \notag\\
\pi_r &= \pi^{\ast} \cdot \mathbf{P} \cdot P_0 \cdot \dotso \cdot P_{r-1} \notag\\
\pi_r &= \pi^{\ast} \cdot P_0 \cdot \dotso \cdot P_{r-1} \cdot P_r \cdot P_{r+1} \cdot \dotso \cdot P_{T-1} \cdot P_0 \cdot \dotso \cdot P_{r-1} \notag \\
\pi_r &= \pi_r \cdot (P_{r} \cdot P_{r+1} \cdot \dotso \cdot P_{T-1} \cdot P_0 \cdot P_1 \cdot \dotso \cdot P_{r-1}). \label{balanceeq}
\end{align}

\subsection{Persistence}

The persistence of a given state $s_i$ is related with the number of steps the system consecutively remains in this state. In the time-homogeneous case, it follows a geometric distribution with expected value $(1-p_{i,i})^{-1}$ and is denoted by $\tau$\nomenclature{$\tau$}{Persistence}. \citet{Anastasiou1996} showed that it is possible to determine the expected time that a Markov chain stays consecutively inside a given subset of states using a simple closed-form expression. For example, in wind power applications, a typical subset of interest could contain all states corresponding to power production above a given threshold. To compute this estimate, the states are renumbered, s.th. they can be partitioned into two disjoint subsets: $\mathcal{A} = \{s_{\nu},...,s_{n}\}$\nomenclature{$\mathcal{A}$}{Subset of the state space, containing the states of interest for persistence} containing the states of interest; and $\overline{\mathcal{A}} = \{s_0,...,s_{\nu-1}\}$, its complement. 
Then, the transition matrix is rearranged into the following block structure:

\begin{equation*}
P = 
\begin{pmatrix}
A & {B} \\ 
{C} & {D}
\end{pmatrix}
=
\begin{pmat}[{..|}]
p_{0,0} & \cdots & p_{0, \nu-1} & p_{0, \nu} & \cdots & p_{0, n} \cr
\vdots & & \vdots & \vdots & & \vdots \cr
p_{\nu-1,0} & \cdots & p_{\nu-1, \nu-1} & p_{\nu-1, \nu} & \cdots & p_{\nu-1, n} \cr\-
p_{\nu,0} & \cdots & p_{\nu, \nu-1} & p_{\nu, \nu} & \cdots & p_{\nu, n} \cr 
\vdots & & \vdots & \vdots & & \vdots \cr
p_{n,0} & \cdots & p_{n, \nu-1} & p_{n, \nu} & \cdots & p_{n, n} \cr
\end{pmat},
\label{rearrangecont}
\end{equation*}

where the first and last block of rows and columns correspond to the states in subset $\overline{\mathcal{A}}$ and $\mathcal{A}$, respectively. The expected value of persistence, i.e. the expected number of time steps the Markov process consecutively remains in the subset $\mathcal{A}$ once it is entered, is given by:

\begin{equation*}
E\lbrace \tau \rbrace = \frac{\pi_{\mathcal{A}} \mathbf{1}_\mathcal{A}}{{\pi}_{\mathcal{A}} {C} \mathbf{1}_{\mathcal{B}}},
\label{eq_persistence}
\end{equation*}

\nomenclature{$E[$ $]$}{Expected value operator}where ${\pi}_{\mathcal{A}}$\nomenclature{${\pi}_{\mathcal{A}}$}{Stationary probability distribution of the states in subset $\mathcal{A}$} is the stationary probability distribution of the states in subset $\mathcal{A}$ and $\mathbf{1}_\mathcal{A}$\nomenclature{$\mathbf{1}_\mathcal{A}$}{unit column vector of the same size as subset $\mathcal{A}$} is the unit column vector of size $(n - \nu + 1) \times 1$ \citep{Anastasiou1996}.\\
For the time-inhomogeneous case, persistence $\tau_t$ is defined as the number of time steps the Markov process is expected to remain in a state (set of states), once it is entered at time $t$. For a cyclic Markov process, the persistence $\tau_t$ is equal for all $t$ that are congruent modulo $T$. Thus, it is only necessary to compute the persistence for $\tau_r$,~$r = 0,...,T-1$\nomenclature{$\tau_r$}{Time-dependent persistence in a cyclic Markov process}. This can be achieved by adapting the derivation of equation~\ref{eq_persistence}, provided by \citet{Anastasiou1996}, to time-variant cyclic Markov processes.\\
After renaming, s.th. the subset of interest is $\mathcal{A} = \{s_{\nu},...,s_{n}\}$, the states of each transition matrix $P_r$ are rearranged as in equation~\ref{rearrangecont}.

\begin{equation*}
P_r = 
\begin{pmatrix}
{A_r} & {B_r} \\ 
{C_r} & {D_r}
\end{pmatrix}
=
\begin{pmat}[{..|}]
p_{0,0}(r) & \cdots & p_{0, \nu-1}(r) & p_{0, \nu}(r) & \cdots & p_{0, n}(r) \cr
\vdots & & \vdots & \vdots & & \vdots \cr
p_{\nu-1,0}(r) & \cdots & p_{\nu-1, \nu-1}(r) & p_{\nu-1, \nu}(r) & \cdots & p_{\nu-1, n}(r) \cr\-
p_{\nu,0}(r) & \cdots & p_{\nu, \nu-1}(r) & p_{\nu, \nu}(r) & \cdots & p_{\nu, n}(r) \cr 
\vdots & & \vdots & \vdots & & \vdots \cr
p_{n,0}(r) & \cdots & p_{n, \nu-1}(r) & p_{n, \nu}(r) & \cdots & p_{n, n}(r) \cr
\end{pmat},
\end{equation*}

The probability of $\tau_r$ to be equal to $l$ is given as:

\begin{align}
Pr(\tau_r = l) &= Pr(X_r \in \mathcal{A},...,X_{r + l} \in \mathcal{A},X_{r+l+1}\notin \mathcal{A} \mid X_r \in \mathcal{A}, X_{r-1} \notin \mathcal{A}) \notag\\
&= \sum_{i \in \mathcal{A}}Pr(X_r = i \mid X_r \in \mathcal{A}, X_{r-1} \notin \mathcal{A}) \cdot Pr(X_s \in \mathcal{A}, r < s \leq l, X_{r+l+1} \notin \mathcal{A} \mid X_r = i)\notag\\
&= \sum_{i \in \mathcal{A}}\sum_{k \notin \mathcal{A}} \sum_{j \in \mathcal{A}} \tilde{\pi}_r(i) \cdot \mathcal{P}_{i,j}(r, l-1, \mathcal{A}) \cdot p_{j,k}(r+l)\label{probpersistence}
\end{align}

with 

\begin{align*}        
\mathcal{P}_{i,j}(r, l, \mathcal{A}) &=Pr(X_{r+l} = j, X_{k} \in \mathcal{A}, 0 < k < l \mid X_r = i)\\
&= D_{r} \cdot \dotso \cdot D_{r + l - 1} \cdot C_{r + l} \cdot \mathbf{1}_{\overline{\mathcal{A}}} 
\end{align*}

and

\begin{align}
\tilde{\pi}_r(i) &= Pr(X_r = i \mid X_r \in \mathcal{A}, X_{r-1} \notin \mathcal{A}) \notag \\
&= \frac{\sum_{j \notin \mathcal{A}}Pr(X_r = i \mid X_{r-1} = j) \cdot Pr(X_{r-1} = j)}{\sum_{i \in \mathcal{A}} \sum_{j \notin \mathcal{A}} Pr(X_r = i \mid X_{r-1} = j) \cdot Pr(X_{r-1} = j)} \notag \\
&= \frac{\sum_{j \notin \mathcal{A}} \pi_{r-1}(j)p_{j,i}(r-1)}{\sum_{k \in \mathcal{A}} \sum_{j \notin \mathcal{A}} \pi_{r-1}(j)p_{j,k}(r-1)}, \quad i \in \mathcal{A},
\label{eq_stat}
\end{align}

where $\pi_r(j)$\nomenclature{$\pi_r(j)$}{Stationary probability, of state $j$ at time of the day $r$} is the long term probability of occurrence (stationary probability) of state $j$ at time of the day $r$; also note that $\pi_t(j) = \pi_r(j)$ for $t=mr$, $\forall m \in \mathbb{N}$. Equation \ref{eq_stat} can be rewritten in the matrix form to include all states in the subset $\mathcal{A}$:

\begin{equation*}
\tilde{\pi}_r(\mathcal{A}) = \frac{\pi_{r-1}({\overline{\mathcal{A}}}) \cdot B_{r-1}}{\pi_{r-1}(\overline{\mathcal{A}}) \cdot B_{r-1} \cdot \mathbf{1}_{\mathcal{A}}},
\end{equation*}

where $\mathbf{1}_{\mathcal{A}}$ is a unit vector of dimension $(n - \nu + 1) \times 1$ and $\pi_{r-1}(\overline{\mathcal{A}})$\nomenclature{$\pi_{r}(\mathcal{A})$}{Vector whose elements are the stationary probabilities of the states in the set $\mathcal{A}$ at time of the day $r$} is a vector of dimensions $1 \times \nu$, whose elements are the stationary probabilities of the states in the set $\overline{\mathcal{A}}$ at time of the day $r-1$.            

Thus, equation~\ref{probpersistence} can be rewritten as:

\begin{align*}
Pr(\tau_r = l) &= \tilde{\pi}_r({\mathcal{A}}) \cdot D_{r} \cdot \dotso \cdot D_{r + l - 1} \cdot C_{r + l} \cdot \mathbf{1}_{\overline{\mathcal{A}}}\\
&= \frac{\pi_{r-1}({\overline{\mathcal{A}}}) \cdot B_{r-1}}{\pi_{r-1}({\overline{\mathcal{A}}}) \cdot B_{r-1} \cdot \mathbf{1}_{\mathcal{A}}} \cdot D_{r} \cdot \dotso \cdot D_{r + l - 1} \cdot C_{r + l} \cdot \mathbf{1}_{\overline{\mathcal{A}}}.
\end{align*}

The expected value of persistence at time $r$ can then be derived as:

\begin{equation*}
E(\tau_r) = \sum_{l = 1}^{\infty} l \cdot \frac{\pi_{r-1}({\overline{\mathcal{A}}}) \cdot B_{r-1}}{\pi_{r-1}({\overline{\mathcal{A}}}) \cdot B_{r-1} \cdot \mathbf{1}_{\mathcal{A}}} \cdot D_{r} \cdot \dotso \cdot D_{r + l - 1} \cdot C_{r + l} \cdot \mathbf{1}_{\overline{\mathcal{A}}}\\
\end{equation*}

Making use of the cyclicity of the Markov process, this can be expressed as:

\begin{equation*}
E(\tau_r) = \sum_{l = 1}^{\infty} l \cdot \frac{\pi_{r-1}({\overline{\mathcal{A}}}) \cdot B_{r-1}}{\pi_{r-1}({\overline{\mathcal{A}}}) \cdot B_{r-1} \cdot \mathbf{1}_{\mathcal{A}}} \cdot \mathcal{D}^{m} \cdot D_{r} \cdot \dotso \cdot D_{r + r_l - 1} \cdot C_{r + r_l} \cdot \mathbf{1}_{\overline{\mathcal{A}}}
\end{equation*}

where $\mathcal{D} = D_{r} \cdot \dotso \cdot D_T \cdot D_{T+1} \cdot \dotso \cdot D_{r-1}$ and $l = mT + r_l$.\\

It can be seen that the sum converges after splitting it into $T$ partial sums, one for each time of the day $r$. For each partial sum, the only term not constant is the matrix power $\mathcal{D}^m$, which converges because all $\mathcal{D}$ eigenvalues are smaller than $1$. The infinite sum for the expected value of persistence at time $r$ can be approximated to an arbitrary degree of accuracy $\epsilon$ by defining

\begin{equation*}
f_l = l \cdot \frac{\pi_{r-1}({\overline{\mathcal{A}}}) \cdot B_{r-1}}{\pi_{r-1}({\overline{\mathcal{A}}}) \cdot B_{r-1} \cdot \mathbf{1}_{\mathcal{A}}} \cdot \mathcal{D}^{m} \cdot D_{r} \cdot \dotso \cdot D_{r + r_l - 1} \cdot C_{r + r_l} \cdot \mathbf{1}_{\overline{\mathcal{A}}}.
\end{equation*}

and successively adding $f_l$,~$l = 0,1,...,L$ until the difference between two consecutive sums is smaller than $\epsilon$, i.e. until $\mid f_L \mid < \epsilon$.

\section{Parameter estimation}
\label{ParameterEstimation}

\subsection{Time-homogeneous Markov chain}
\label{timehomogeneous}

The common approach to estimate the Markov chain transition matrix $P$ is through the optimization of a constrained maximum likelihood function, which describes the realization probability of a given dataset \citep{Anderson1957}. For a sequence of $M$ states $\lbrace X_0 = s_{i_0}, ... , X_M = s_{i_M} \rbrace$ with $s_{i_0}, ..., s_{i_M} \in S$ and $i_0, ..., i_M \in \{0,...,n\}$, its probability can be computed as $Pr\{X_0 = s_{i_0}\}p_{i_0,i_1} p_{i_1,i_2} \cdot \ldots \cdot p_{i_{M-1},i_M}$. Since the term $Pr\{X_0 = s_{i_0}\}$ is constant, given a set of observed state transitions $\mathcal{S}$\nomenclature{$\mathcal{S}$}{Set of observed state transitions}, it is possible to estimate ${P}$ by maximizing the likelihood

\begin{equation}
OF_1 = \prod_{(i,j)\in \mathcal{S}} p_{i,j},
\label{OF1}
\end{equation}

where a transition is described by an ordered pair $(i,j)$ indicating the origin and the destination of the transition. 
In practice, instead of maximizing $OF_1$ with respect to the $p_{i,j}$ variables it is preferable to minimize the negative log-likelihood function, i.e. $- \log(OF_1)$, since it transforms the original mathematical programming problem into an equivalent one that is convex and, thus, has a unique solution \citep{Boyd2004}.
The overall optimization problem is formulated as follows:

\begin{align*}
& \text{min}
& & - \sum_{(i,j)\in \mathcal{S}} \log(p_{i,j})\\
& \text{subject to}
& &  p_{i,j} \geq 0 \quad \forall i,j = 0,...,n\\
& & & \sum_{j}{p_{i,j}} = 1 \quad \forall i = 0,...,n
\end{align*}
The constraints ensure non-negativity of the transition probabilities and that they sum up to $1$ for each row of the transition matrix. 

\subsection{Cyclic time-variant Markov process}
\label{timevariant}

The goal of this time-variant Markov process is to get a model that accurately reproduces the long-term behavior while considering the daily patterns observed in the data. Thus, the proposed objective function combines two maximum likelihood estimators: the first term maximizes the likelihood of the cycle-average probability; and, the second term maximizes the likelihood of the time-dependent probability. The final optimization problem is transformed into a convex one using the negative logarithm of the objective function.
This section provides a detailed description of the objective function, the parametrization of the time-variant probability functions, and the constraints that must be added to the optimization problem to ensure its Markov properties.

\subsubsection{Objective function}

The transition probabilities are considered to be time-variant and cyclic with a period of $T$, i.e. for each time of the day $r$~($=0,...,T-1$) there is a different transition matrix ${P_r}$.
In this paper, the time-dependent transition probabilities $p_{i,j}(r)$ are modeled by Bernstein polynomials. This has several advantages: a) a polynomial representation of the transition probabilities leads to a convex objective function and constraints, i.e. the optimization problem has a unique solution; b) a polynomial representation allows to decrease the number of variables in the optimization problem: for each transition, instead of $T$ variables only $k+1$ are needed for a $k$ order polynomial; c) Bernstein polynomials are non-negative, which simplifies probability modeling, when compared to other polynomial bases; and d) they have the convex hull property, which, combined with de Casteljau algorithm, allows to easily write probability boundary conditions.\\
Bernstein polynomials are linear combinations of Bernstein basis polynomials $b_{\mu, k}(z)$, $z \in [0,1]$\nomenclature{$b_{\mu, k}(z)$}{$\mu$-th Bernstein basis polynomial of order $k$}. The $k + 1$ Bernstein basis polynomials of order $k$ are defined as:

\begin{equation*}
b_{\mu, k}(z) = \left(\! \begin{array}{c} k \\ \mu \end{array} \!\right) z^{\mu} (1-z)^{k - \mu}
\end{equation*}

with $\mu = 0, ..., k$ and $\left(\! \begin{array}{c} k \\ \mu \end{array} \!\right)$ the binomial coefficient. Thus, the transition probabilities $p_{i,j}(z)$ are described by

\begin{equation*}
p_{i,j}(z) = \sum_{\mu = 0}^k \beta_{\mu}^{i,j}b_{\mu,k}(z),
\end{equation*}

with $\beta_{\mu}^{i,j} \in \mathbb{R}$\nomenclature{$\beta_{\mu}^{i,j}$}{Coefficients of the Bernstein polynomial modeling the transition probability $p_{i,j}(t)$} and $z = \frac{r}{T}$\nomenclature{$z$}{Scaled time of the day}, since the polynomial variable has to be scaled, s.th. it is between 0 and 1.\\

Hence, to maximize the likelihood of the time-dependent transition probabilities given the data, the objective function must consider the time of the day $z$ when the transition happens. Therefore, the objective function introduced in~(\ref{OF1}) becomes $\sum_{(i,j)_z\in \mathcal{S}_z} \log(p_{i,j}(z))$, where $\mathcal{S}_z$\nomenclature{$\mathcal{S}_z$}{Set of transitions observed in the data together with the scaled time of the day $z$ at which they are observed} is the set of observed transitions together with the time $z$ when they happens. This objective function allows to compute the intra-cycle transition probability functions, and thus to represent the daily patterns present in the data.

A second term is added to this function, namely $\sum_{(i,j)\in \mathcal{S}} \log(p^{avg}_{i,j})$, where $\mathcal{S}$ is the set of transitions observed in the data as defined in section~\ref{timehomogeneous} and $p_{i,j}^{avg}$\nomenclature{$p_{i,j}^{avg}$}{Daily average probability of transition from state $s_i$ to $s_j$} is the cycle-average (daily) probability of transition from state $s_i$ to $s_j$. It can be computed as follows:



\begin{align*}
p_{i,j}^{avg} &= \frac{1}{1-0}\int_0^1p_{ij}(z)dz = \int_0^1\sum_{\mu = 0}^k\beta^{i,j}_kb_{\mu,k}(z)dz\\
&= \sum_{\mu = 0}^k \beta^{i,j}_k \int_0^1b_{\mu,k}(z)dz = \frac{1}{k + 1} \sum_{\mu = 0}^k \beta^{i,j}_k
\end{align*}

This second term is the maximum likelihood estimator for the daily average probability and its addition to the objective function increases the consistency of the long-term behavior of the Markov process with the data. Therefore, the overall objective function $OF_2$ is given by:

\begin{align*}
OF_2 &= 
- \sum_{(i,j)\in \mathcal{S}} \log(p^{avg}_{i,j}) - \sum_{(i,j)_z\in \mathcal{S}_z} \log(p_{i,j}(z))\\
&= - \sum_{(i,j)\in \mathcal{S}} \log(\frac{1}{k + 1} \sum_{\mu = 0}^k \beta^{i,j}_k) - \sum_{(i,j)_z\in \mathcal{S}_z} \log(\sum_{\mu = 0}^k \beta_{\mu}^{i,j}b_{\mu,k}(z))
\end{align*}

\noindent and minimization is performed with respect to the coefficients $\beta^{i,j}_{\mu}$ (model parameters).

\subsubsection{Constraints}

The estimation of the model parameters requires the transition probability functions to comply with several constraints, to ensure:
\begin{itemize}
\item $\mathcal{C}^0$- and $\mathcal{C}^1$-continuity at $z = 0$,
\item row-stochasticity of the transition matrices at every time of the day $z$ and
\item that the transition probability functions are non-negative and bounded by 1.
\end{itemize}
Thus, to complete the specification of the optimization problem this section explains all the necessary constraints required for the model parameters to describe a cyclic Markov process.

\textit{Periodicity}

The transition probability functions are modeled using Bernstein polynomials, which are smooth, i.e. $\mathcal{C}^{\infty}$-continuous functions. In general, the values at both ends of their domain ($0$ and $1$) need not be equal. Thus, to avoid sudden changes in the value and slope of each probability function between the cycles, two constraints are added to ensure $\mathcal{C}^0$ and $\mathcal{C}^1$-continuity. Another reason is the arbitrariness of the cycle starting position, which affects the position of the discontinuity if these conditions are not used.

The first constraint is $p_{i,j}(0) = p_{i,j}(1)$. Since $b_{\mu, k}(0) = \delta_{\mu, 0}$ and $b_{\mu, k}(1) = \delta_{\mu, k}$ the constraint can be reformulated as $\beta_0^{i,j} = \beta_k^{i,j}$, where $\delta$ is the Kronecker delta. The second constraint is added to ensure $\mathcal{C}^1$-continuity, i.e. $\frac{d p_{i,j}}{dz}(0) = \frac{d p_{i,j}}{dz}(1)$. The first derivative of a Bernstein basis polynomial can be written as a combination of two polynomials of lower degree:

\begin{equation*}
\frac{d b_{\mu, k}}{dz}(z) = k(b_{\mu - 1, k - 1}(z) - b_{\mu, k - 1}(z))
\end{equation*}

Thus, the first derivative of a transition probability $p_{i,j}(z)$ is given by:

\begin{equation*}
\frac{d p_{i,j}}{dz}(z) = k (\sum_{\mu = 1}^k (\beta_{\mu}^{i,j} - \beta_{\mu - 1}^{i,j}) b_{\mu -1, k-1}(z) - \beta_k^{i,j} b_{k, k-1}(z))
\end{equation*}


Hence, using $b_{\mu, k}(0) = \delta_{\mu, 0}$ and $b_{\mu, k}(1) = \delta_{\mu, k}$ as well as the first constraint $\beta_0^{i,j} = \beta_k^{i,j}$~$\forall i,j = 0,...,n$, the constraint $\frac{d p_{i,j}}{dz}(0) = \frac{d p_{i,j}}{dz}(1)$ reduces to the following linear constraint:

\begin{equation*}
\beta_{k}^{i,j} = \beta_{0}^{i,j} = 0.5(\beta_{1}^{i,j} + \beta_{k-1}^{i,j})
\end{equation*}


\textit{Row stochasticity of transition matrices}

To ensure row stochasticity of the time-variant transition matrices, it is necessary to ensure that $\sum_{j} p_{i,j}(z) = 1$ for all $i$ and $z$. Since the Bernstein basis polynomials of order $k$ form a partition of unity, i.e.

\begin{equation*}
\sum_{\mu = 0}^k b_{\mu, k}(z) = 1
\end{equation*}


the constraint can be re-written as a linear combination of the polynomial coefficients:


\begin{align*}
\sum_{j} p_{i,j}(z) = 1 & \Leftrightarrow \sum_{j} \sum_{\mu = 0}^k \beta_{\mu}^{i,j}b_{\mu,k}(z) = 1\\
& \Leftrightarrow \sum_{j} \beta_{\mu}^{i,j} = 1\\
\end{align*}

\textit{Non-negative transition probabilities are bounded by 1}

The most straightforward way to implement this constraint is to add two inequalities for each time of the day and each transition probability $p_{i,j}$, i.e.

\begin{equation}
\begin{aligned}
p_{i,j}(z) &\geq 0 \quad \forall i,j,z\\
p_{i,j}(z) &\leq 1 \quad \forall i,j,z
\label{bound_constraint_original}
\end{aligned}
\end{equation}

However, this constraint significantly increases the problem size, since it requires $2 \cdot T \cdot n^2$ inequalities. An alternative constraint can be formulated by using the convex hull property of the Bernstein polynomials. This constraint makes the overall optimization problem size smaller, but is more restrictive. 

Every Bernstein polynomial $\sum_{\mu = 0}^k \beta_{\mu}b_{\mu,k}(z)$ always lies in the convex hull defined by its control points $ (\frac{k}{\mu}, \beta_{\mu})$,~$\mu = 0,...,k$. Thus the constraint

\begin{equation*}\label{probconstraint}
0 \leq p_{i,j}(z) \leq 1 \quad \forall i,j = 0,...,n
\end{equation*}

can be reformulated in terms of the polynomial coefficients as

\begin{equation}\label{coeffconstraint}
0 \leq \beta_{\mu}^{i,j} \leq 1 \quad \forall \mu = 0,...,k
\end{equation}

Since constraint~\ref{coeffconstraint} is a sufficient but not necessary condition for constraint~\ref{bound_constraint_original}, the reformulation leads to a more restrictive overall minimization problem, i.e. the optimum objective function value is always higher or equal when compared with the problem with original constraint~\ref{bound_constraint_original}.
The convex hull bound of Berstein polynomials can be tightened by subdivision, i.e. by subdividing the domain in two regions and finding new control points $\beta^{i,j}_0(1),...,\beta_k^{i,j}(1)$ and ${\beta^{i,j}_{k+1}}(1),...,\beta^{i,j}_{2k}(1)$ such that the function output remains unchanged. With each subdivision, the control points form a tighter bound around the polynomial and thus the polynomial coefficients can assume values in a wider range. The new control points represent the polynomial restricted to the two sub-intervals $[0, z^{\ast}]$ and $[z^{\ast}, 1]$, where $z^{\ast} \in [0,1]$ is the cutting point of the division. For simplicity, $z^{\ast}$ is fixed to 0.5 in all transition probabilities.
The new control points can be determined by linear combinations of the original control points $\beta^{i,j}_0,...,\beta^{i,j}_k$. This can be performed efficiently using de Casteljau algorithm, which in matrix form is given as:

\begin{equation}
\begin{pmatrix}
\beta^{i,j}_0(1)\\
\vdots \\
\beta^{i,j}_k(1)
\end{pmatrix}
=
\begin{pmatrix}
b_{0,0}(z^{\ast}) & 0 & \cdots & 0 \\
b_{0,1}(z^{\ast}) & b_{1,1}(z^{\ast}) & \cdots & 0 \\
\vdots  & \vdots  & \ddots & \vdots  \\
b_{0,k}(z^{\ast}) & b_{1,k}(z^{\ast}) & \cdots & b_{k,k}(z^{\ast})
\end{pmatrix}
\begin{pmatrix}
\beta^{i,j}_0\\
\vdots \\
\beta^{i,j}_k
\end{pmatrix}
= C_l 
\begin{pmatrix}
\beta^{i,j}_0\\
\vdots \\
\beta^{i,j}_k
\end{pmatrix}
= C_l \cdot \beta^{ij}
\label{Casteljauleft}
\end{equation}

and

\begin{equation}
\begin{pmatrix}
\beta^{i,j}_{k+1}(1)\\
\vdots \\
\beta^{i,j}_{2k}(1)
\end{pmatrix}
=
\begin{pmatrix}
b_{0,k}(z^{\ast}) & b_{1,k}(z^{\ast}) & \cdots & b_{k,k}(z^{\ast}) \\
0 & b_{0,k-1}(z^{\ast}) & \cdots & b_{k-1,k-1}(z^{\ast}) \\
\vdots  & \vdots  & \ddots & \vdots  \\
0 & \cdots & 0 & b_{0,0}(z^{\ast}) \\
\end{pmatrix}
\begin{pmatrix}
\beta^{i,j}_0\\
\vdots \\
\beta^{i,j}_k
\end{pmatrix}
= C_r 
\begin{pmatrix}
\beta^{i,j}_0\\
\vdots \\
\beta^{i,j}_k
\end{pmatrix}
= C_r \cdot \beta^{ij}
\label{Casteljauright}
\end{equation}

The subdivision can be applied recursively to further improve the convex bound around the polynomial. The corresponding coefficients are computed by applying equations~\ref{Casteljauleft} and~\ref{Casteljauright} to the left and right coefficient vectors. Defining $C = (C_l, C_r)^T$ and $I_z$ as the identity matrix of dimension $z \times z$, the coefficients $\beta^{i,j}(w) = (\beta^{i,j}_0(w),...,\beta^{ij}_{2^{w}k}(w))$ after $w$ subdivisions can be obtained by:

\begin{equation*}
\beta^{i,j}(w) = (C \otimes I_{2^{w-1}}) \cdot (C \otimes I_{2^{w-2}}) \cdot \dotso \cdot (C \otimes I_{2}) \cdot (C \otimes I_1) \cdot \beta^{i,j}
\end{equation*}

where $\otimes$ denotes the Kronecker product. The number of inequalities needed for the implementation of this constraint is $(k + 1) \cdot 2^{\omega + 1} \cdot n^2$. Thus, its use only makes sense if it decreases the problem size, i.e. for a number of subdivisions $\omega$ such that $(k + 1) \cdot 2^{\omega + 1} \leq T$.

\subsubsection{Problem formulation}
The overall optimization problem to be solved for the estimation of the transition probability coefficients $\beta_{\mu}^{i,j}$ can be written as:

\begin{align}
& \min_{\beta_{\mu}^{i,j}}
\label{objectivefunction}
& & - \sum_{(i,j)\in \mathcal{S}} \log(\frac{1}{k + 1} \sum_{\mu = 0}^k \beta^{i,j}_{\mu}) - \sum_{(i,j)_z\in \mathcal{S}_z} \log(\sum_{\mu = 0}^k \beta_{\mu}^{i,j}b_{\mu,k}(z))\\
& \text{subject to}
\label{c0}
& & \beta_0^{i,j} = \beta_k^{i,j} \quad \forall i,j = 1,...,n\\
\label{c1}
& & & \beta_{0}^{i,j} = 0.5(\beta_{1}^{i,j} + \beta_{k-1}^{i,j}) \quad \forall i,j = 0,...,n\\
\label{rowstochasticity}
& & & \sum_{j} \beta_{\mu}^{i,j} = 1 \quad \forall i = 0,...,n; \forall \mu = 0,...,k\\
\label{bounds1}
& & & \beta^{i,j}(w) \leq 1 \quad \forall i,j = 0,...,n\\
\label{bounds2}
& & & 0 \leq \beta^{i,j}(w) \quad \forall i,j = 0,...,n
\end{align}

where $w$ is the number of subdivisions and $k$ is the order of the Bernstein polynomials, which have to be specified. The objective function~(\ref{objectivefunction}) is a combination of two negative log-likelihood functions to ensure the Markov process captures both the daily patterns and the long-term behavior of the original data.
The optimization is performed with respect to several constraints: constraints (\ref{c0}) and (\ref{c1}) ensure $\mathcal{C}^0$- and $\mathcal{C}^1$-continuity at $z = 0$. The row-stochasticity of the transition matrix is ensured by constraint~(\ref{rowstochasticity}). The last two constraints~(\ref{bounds1}) and~(\ref{bounds2}) bound the transition probabilities between $0$ and $1$.

It is expected that the objective function decreases with the polynomial order and the number of subdivisions. The parameters of the Markov chain model $\beta_{\mu}^{i,j}$ are estimated by solving the optimization problem using a rigorous numerical solver. The model was formulated making use of the casadi computation framework \citep{Andersson2010} and the optimization was performed by ipopt, a nonlinear interior-point solver \citep{Wachter2006}, which ensures convergence to the global optimum in the case of convex optimization problems.

\section{Application of the cyclic Markov process to wind turbine modeling}
\label{Example}

\subsection{The data}

The data for this study was obtained from a wind power turbine in a wind park located in a mountainous region in Portugal. The time series consists of a three-year period (2009-2011) of historical data gotten from the turbine data logger. The sampling time of 10 minutes leads to 144 samples each day. The data-set comprises three variables, wind power, speed and direction (nacelle orientation). The wind speed information was collected from the anemometer placed in the wind turbine hub. Due to confidentiality, wind power and speed data values are reported as a fraction of the rated power and the cut-out speed, respectively.

\subsection{Markov chain state definition}
\label{Statedefinition}

Discrete Markov chain models require the definition of the states when applied to describe continuous variables. This work proposes to characterize the wind turbine states using three different variables: wind power, speed and direction. As such, each state is defined by all the points inside a polyhedron in three-dimensional space. 

\begin{figure}[H]
\begin{subfigure}[H]{0.5\textwidth}
\centering
\includegraphics[width = \textwidth]{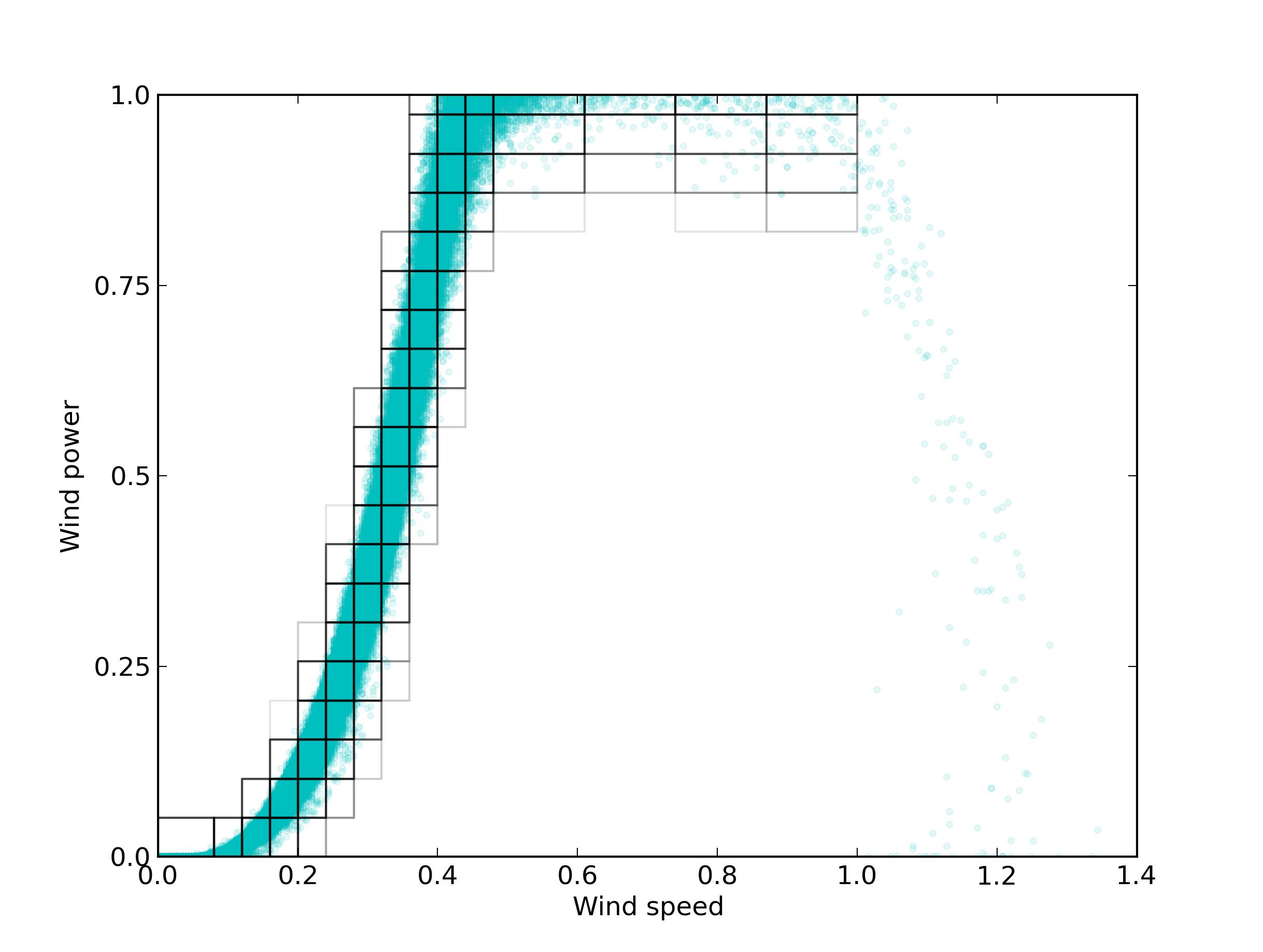}
\end{subfigure}
\begin{subfigure}[H]{0.5\textwidth}
\centering
\includegraphics[width = \textwidth]{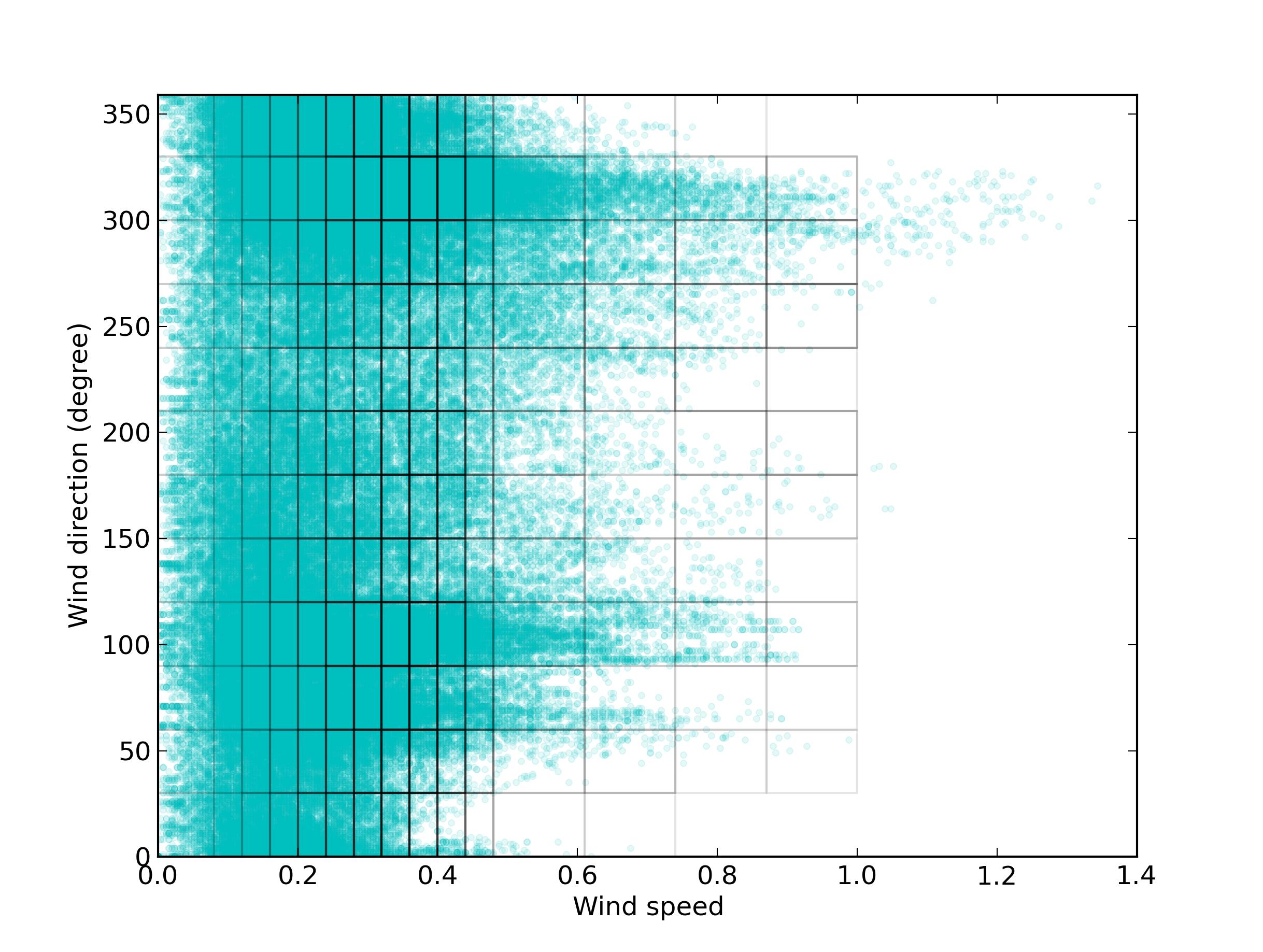}
\end{subfigure}
\caption{Representation of all data points projected into the: a) wind direction and speed plane (left); and, b) wind power and speed plane (right). Each rectangle is the projection of a state polyhedron into the two planes. Overall, they define the final state partition for the three-dimensional variable space.}
\label{fig1}
\end{figure}

Fig.~\ref{fig1} presents all data observations and the state partitions projected into: a) the wind direction and speed plane; and, b) the wind power and speed plane. As expected, the observations projected into the wind power and speed plane define the characteristic power curve of the wind turbine. It shows the three operational regions of a wind turbine: a) below the cut-in speed no power is produced; b) between cut-in and rated wind speed the power increases proportionally to the cube of wind speed; c) at wind speeds between the rated and the cut-out wind speed, the turbine control system limits the power output to a constant value. In the wind direction and speed plane, data is widely scattered and shows the dominant wind patterns at the site. Three accumulation regions can be identified: one for low wind speeds, centered on $0.25$, which is the mode of the wind speed, and two defining the dominant wind directions around 100$^\circ$ and 300$^\circ$.

The data space is discretized unevenly to get a good resolution of the high-slope region of the power curve. In a previous work \citep{Lopes2012}, this partition was used in a time-homogeneous Markov chain and proved to lead to an accurate representation of the original data. The wind direction and power are divided by an equally spaced grid leading to 12 ($\{d_1,...,d_{12}\}$) and 20 ($\{p_1,...,p_{20}\}$) classes, respectively. The wind speed is divided as follows: values below the cut-in speed define one class $sp_1$; between the cut-in and rated wind speed the discretization is narrowed by selecting 10 classes (\{$sp_2,...,sp_{11}\}$); and between the rated and cut-out wind speed discretization is widened and 4 classes ($\{sp_{12},...,sp_{15}\}$) are defined. Data points with wind speed above the cut-out wind speed are discarded. The complete state set is constructed by listing all possible combinations of the classes of each variable. Due to physical constraints between the variables, most of 
the states are empty~(fig.~\ref{fig1}(left)) and can are discarded. This reduces the number of states from 3840 to 778, for this turbine.

\subsection{Additional transitions to promote a single communication class}

The solution of the optimization problem described in section~\ref{ParameterEstimation} comprises a set of transition matrices $P_r$,~$r \in \{0,...,143\}$. However, the constraints in the optimization problem definition do not force the matrix $\mathbf{P} = P_0 \cdot \dotso \cdot P_{143}$ to be irreducible and thus the Markov process to have a single communication class. So, during data synthesis, the Markov process can get ``trapped'' within a communication class. To induce the Markov process to have a single communication class, additional transition counts are introduced into the data. The goal is to add a small set of transitions to promote state connectivity without distorting the original data. Thus, the set is composed of transitions that connect neighboring states in the state space, since those are the ones most likely to occur.\\
For a state $s_i = (p_l, sp_p, d_q)$, its neighborhood $V$ is defined as 

{\footnotesize
\begin{equation}
V(s_i) := \{(p_{ll}, sp_{pp}, d_{qq}): ll \in \{l-1, l, l+1\}, pp \in \{p-1, p, p+1\}, qq \in \{q-1, q, q+1\}\} \setminus s_i.
\end{equation}
}

It should be noted that, unlike power and speed, direction is a circular variable, e.g. states $d_0$ and $d_{11}$ are considered neighbors. If a neighbor state $s_j \in V(s_i)$ is present in the dataset, a transition $(i,j)$ is added to the set of extra transitions $\mathcal{S}_E$. For this dataset, originally consisting of 150601 transitions, 13610 transitions are added.\\
The extra transitions must be considered to happen at an unknown time of the day $z$. Thus, they can only be accounted for in the objective function term without time information, i.e. only in the time-variant part of the objective function. This directly affects the values for $p_{i,j}^{avg}$ and, indirectly, the model parameters. Since the aim is to cause a minimal impact on the transition probabilities, the additional term is weighed by a factor $\omega < 1$\nomenclature{$\omega$}{Weight of the extra transitions added to the objective function} to directly control its influence. In this work it is fixed to 0.05.
Thus, the following term is added to the objective function:

\begin{equation}
- \omega \cdot \sum_{(i,j)\in \mathcal{S}_E} \log(p^{avg}_{i,j}) = - \omega \cdot \sum_{(i,j)\in \mathcal{S}_E} \log(\frac{1}{k + 1} \sum_{\mu = 0}^k \beta^{i,j}_k)
\end{equation}

Although the use of the extra transition set does not ensure the time-variant Markov process to have a single communication class, results show a decrease of the number of communication classes from 13 to 1 in this dataset.

\section{Simulation of wind power, speed and direction time series}
\label{Simulation}
To simulate wind power, speed and direction time series the method described by \citet{Sahin2001} is adapted to the cyclic time-variant Markov model as follows.
First, the cumulative probability transition matrices $P^{cum}_r$ with $P^{cum}_r(i, j) = \sum_{k = 0}^{j}p_{i,k}(r)$ are computed. Then an initial state $s_{i}$, i.e. $X_0 = s_i$, is randomly selected. A new datapoint $X_{t+1}$ is generated by uniformly selecting a random number $\epsilon$ between zero and one. The corresponding state $s_{new}$ ($X_{t+1} = s_{new}$) is chosen such that the probability of reaching it from the current state $s_i$ is bigger than $\epsilon$, i.e. such that $P_{r_t}^{cum}(i, {new}) \geq \epsilon$.\\ 
Based on this discrete state sequence, a real value for the wind power/speed/direction variables is generated by sampling each state partition uniformly.

\section{Results and discussion}
\label{Results}

\subsection{Daily patterns in the data}
\label{dailypatterns}
The wind power, speed and direction time-series clearly show a daily time-dependent behavior.

\begin{figure}[H]
\centering
\begin{subfigure}[H]{0.45\textwidth}
\includegraphics[width = \textwidth]{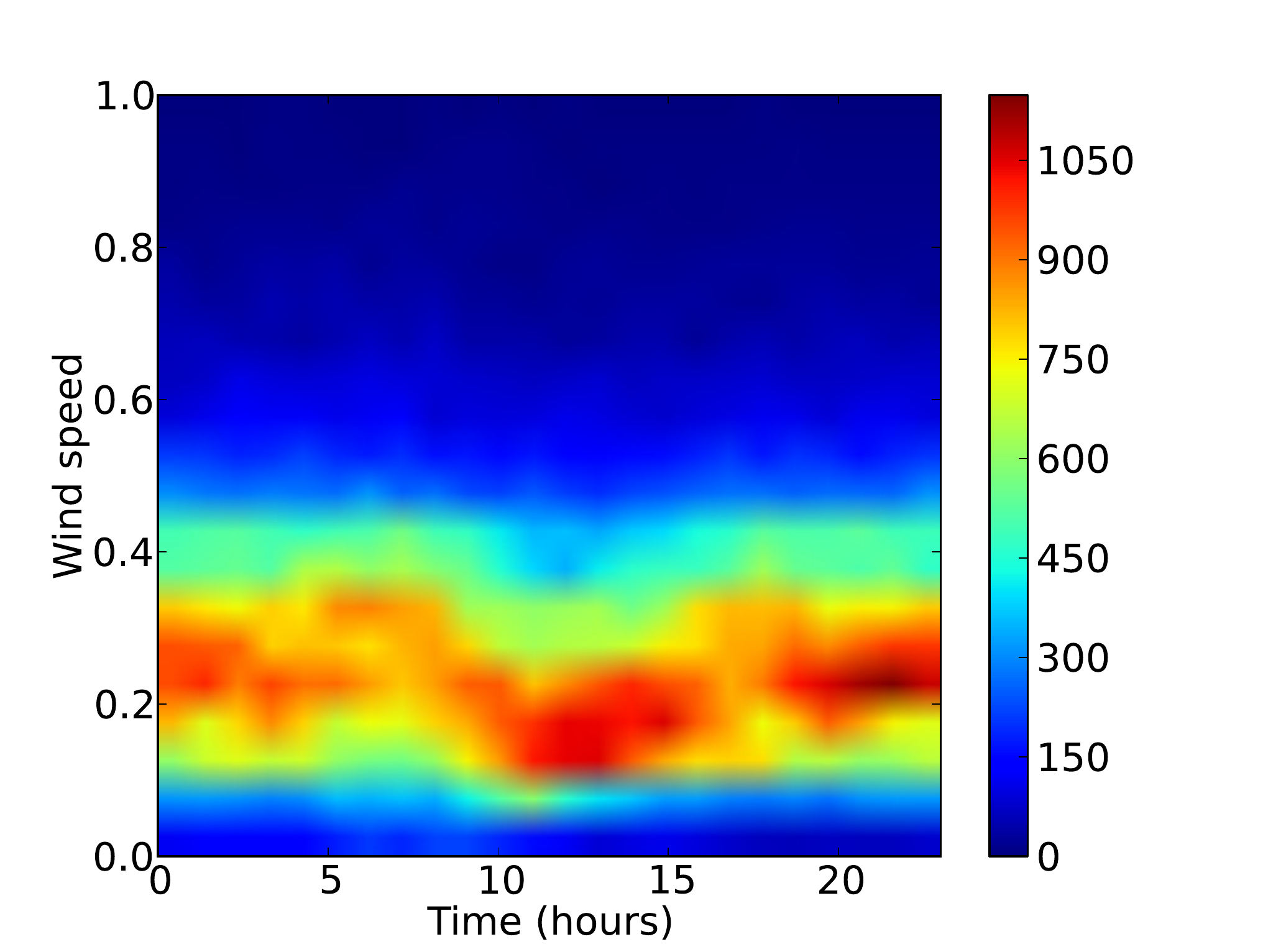}
\end{subfigure}
\begin{subfigure}[H]{0.45\textwidth}
\includegraphics[width = \textwidth]{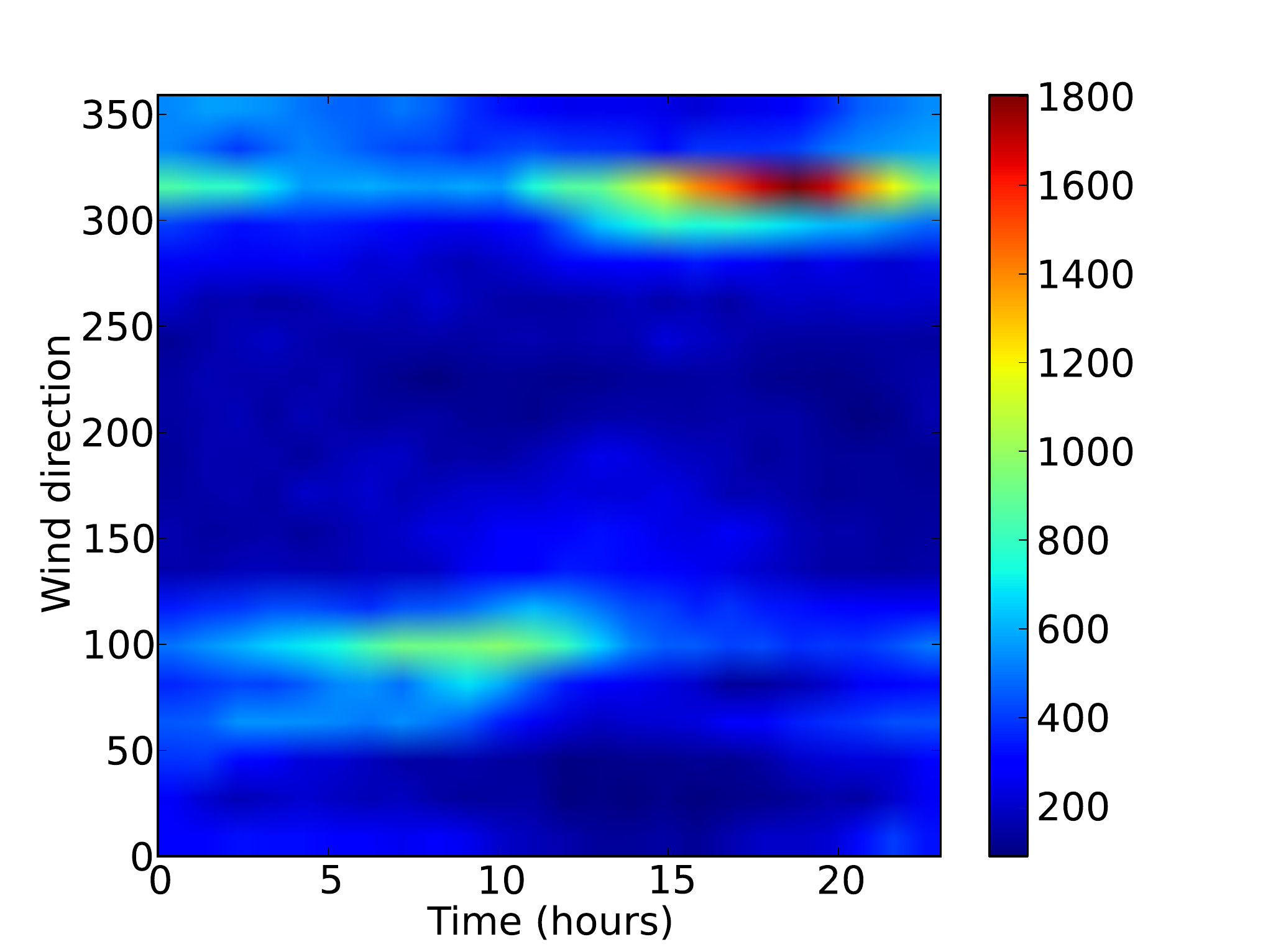}
\end{subfigure}
\begin{subfigure}[H]{0.45\textwidth}
\includegraphics[width = \textwidth]{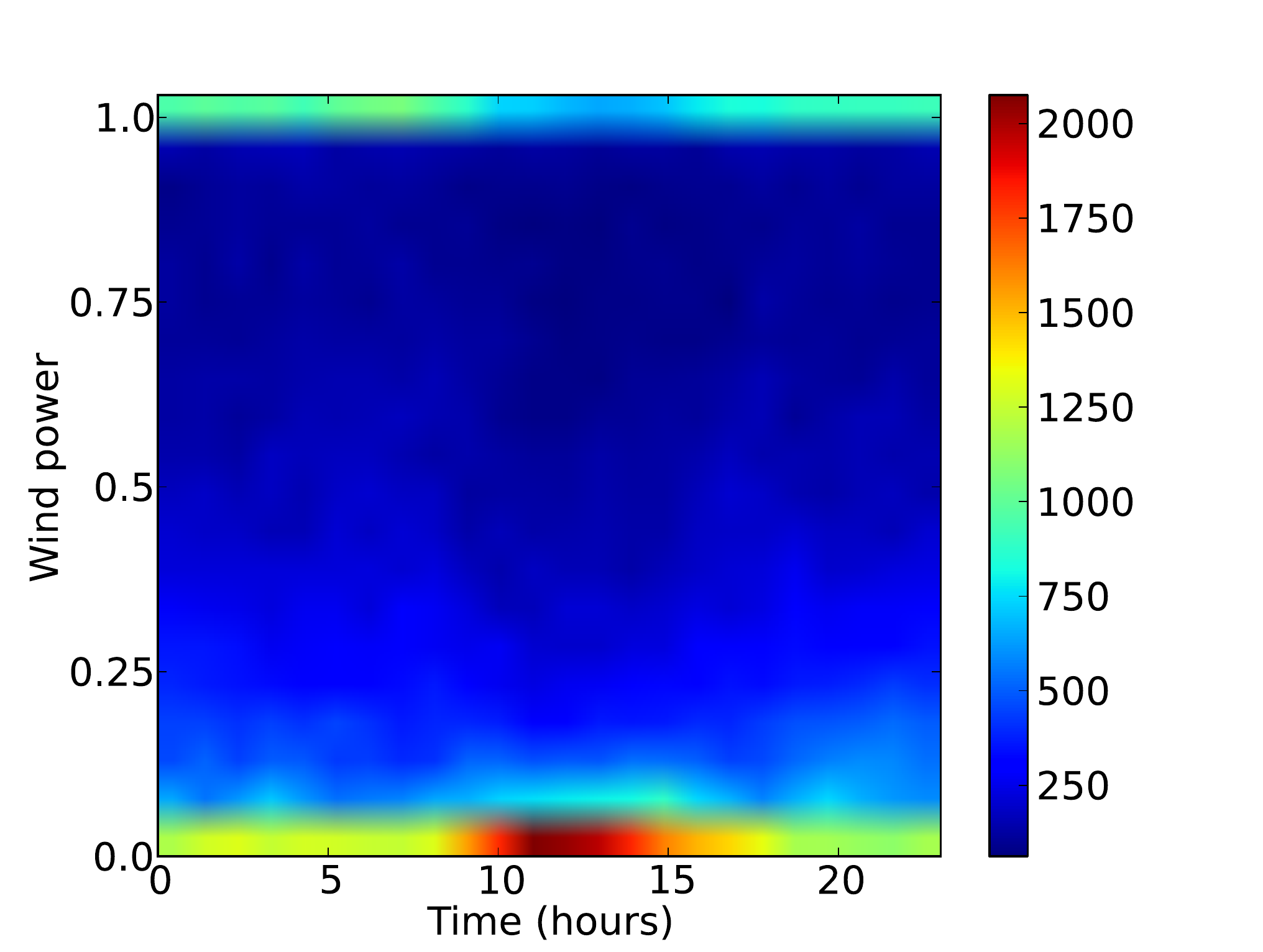}
\end{subfigure}
\begin{subfigure}[H]{0.45\textwidth}
\includegraphics[width = \textwidth]{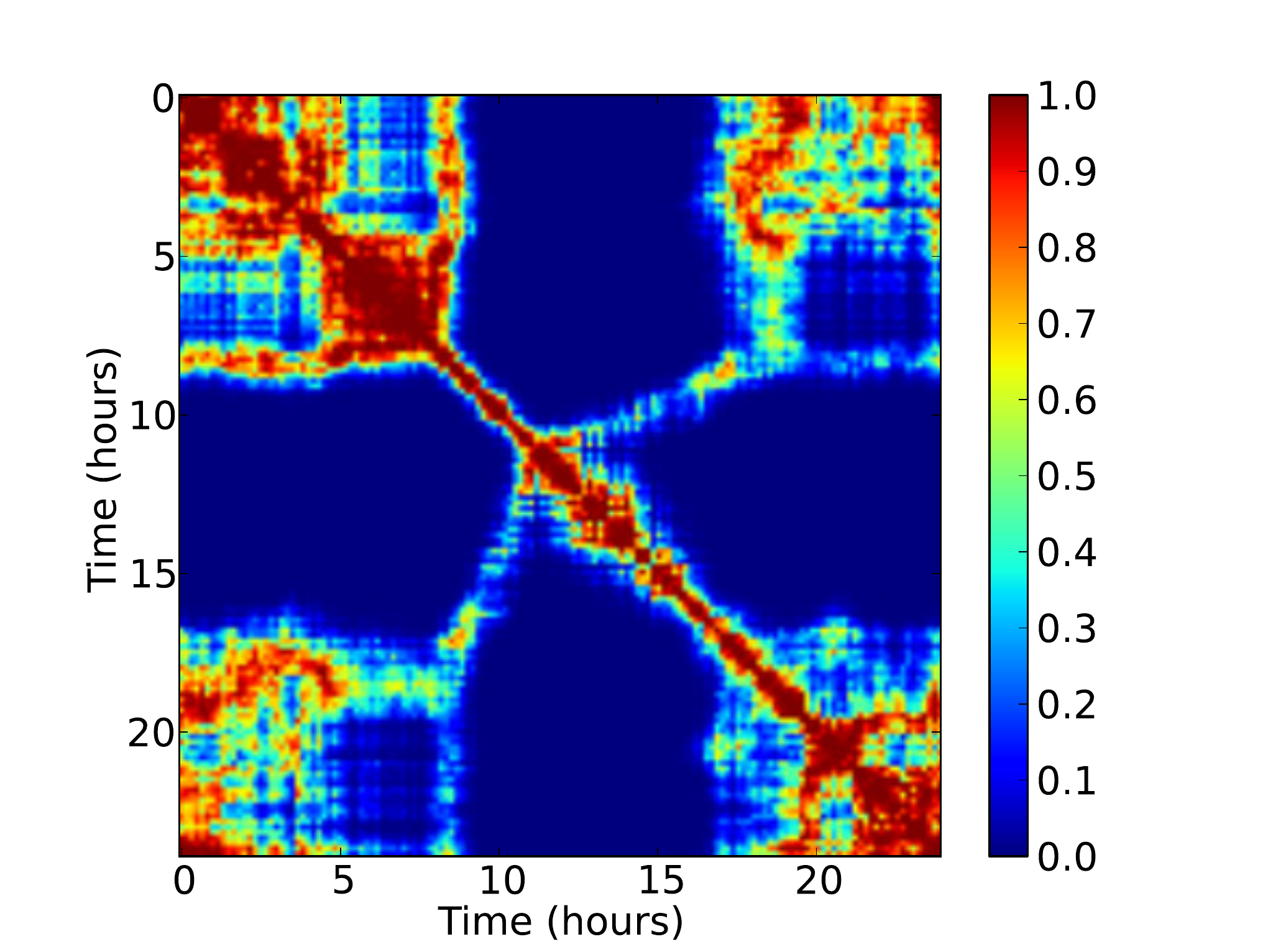}
\end{subfigure}
\caption{Two-dimensional histograms of the original time series data: speed-time (top left), direction-time (top right) and power-time (bottom left). The subfigure on the bottom right shows the p-value of the Kolmogorov-Smirnov test used to compare the wind speed distribution on the different times of the day.}
\label{fig2}
\end{figure}

Figure~\ref{fig2} shows that, on average, the turbine does not produce power between 10 am and 3 pm. In this time interval, low wind speeds (0.1 - 0.25) are the most likely events. There are two dominant wind directions: around 100$^{\circ}$ and 300$^{\circ}$. Moreover, they occur at specific times of the day; between 5 and 10am, the wind typically blows from the 100$^{\circ}$ direction, the rest of the day from 300$^{\circ}$.

To assess whether these two dominant directions might be due to summer/winter seasonality, the dataset was divided in two subsets, one covering the period from April to September and the other from October to March. The histogram analysis shows that both, summer and winter subset, have the same two dominant directions (figures not shown). Thus, it was concluded that the time-dependent pattern is not induced by this seasonality. 

Figure~\ref{fig2} bottom-right plot shows the p-values of the Kolmogorov-Smirnov test applied to the wind speed distributions at different times of the day. The Kolmogorov-Smirnov test is a nonparametric test for the equality of continuous one-dimensional probability distributions. Thus, the high p-values around the diagonal illustrates that wind speed distributions for consecutive times of the day are similar. The same holds for wind speeds in the morning and evening, i.e. before 9am and after 4:30pm. The wind speed distribution between 10am and 3pm is clearly different.

\subsection{Choice of polynomial order and number of subdivisions}

The model introduced in section~\ref{timevariant} has two parameters that need to be defined: $k$, the order of the Bernstein polynomials used to model the transition probabilities; and $w$, the number of subdivisions used to tighten the convex hull that bounds the polynomials. To choose proper values for these parameters, different models were computed by varying $k=4...10$ and $w=0...3$. For each model, synthetic data was generated following the procedure described in section \ref{Simulation} and compared with the real dataset.

\begin{figure}[H]
\centering
\begin{subfigure}[H]{0.45\textwidth}
\includegraphics[width = \textwidth]{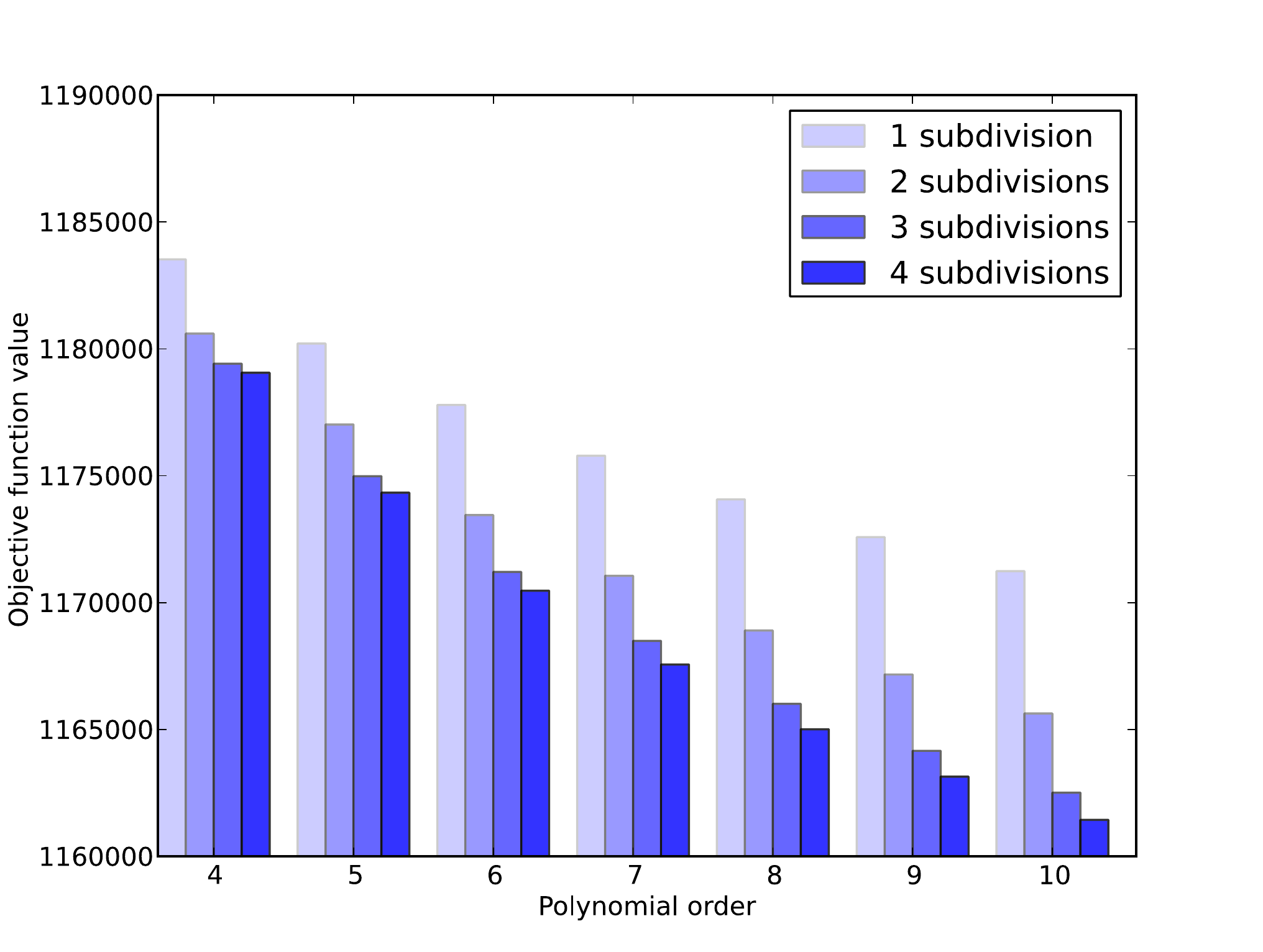}
\end{subfigure}
\begin{subfigure}[H]{0.45\textwidth}
\includegraphics[width = \textwidth]{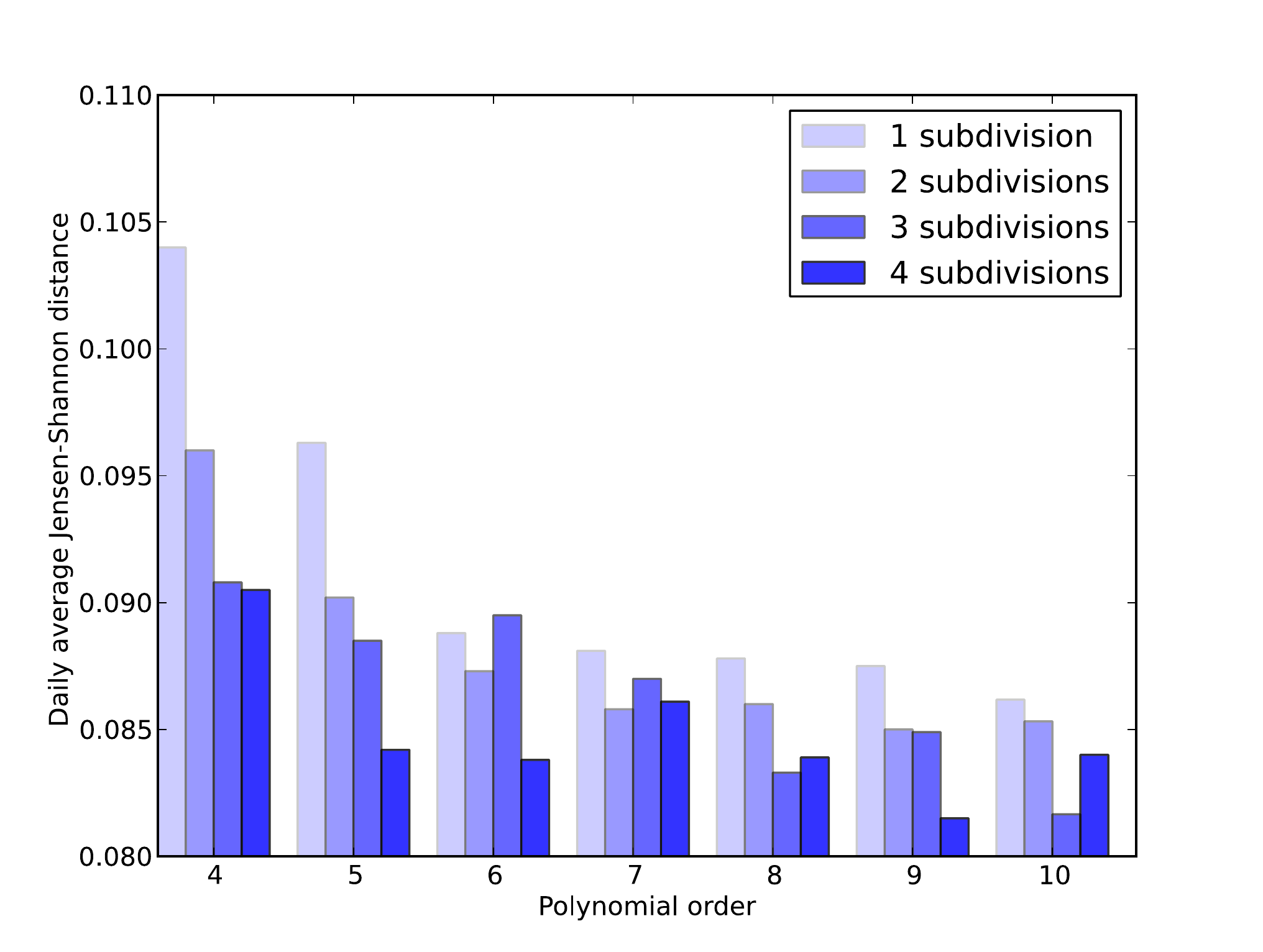}
\end{subfigure}
\begin{subfigure}[H]{0.45\textwidth}
\includegraphics[width = \textwidth]{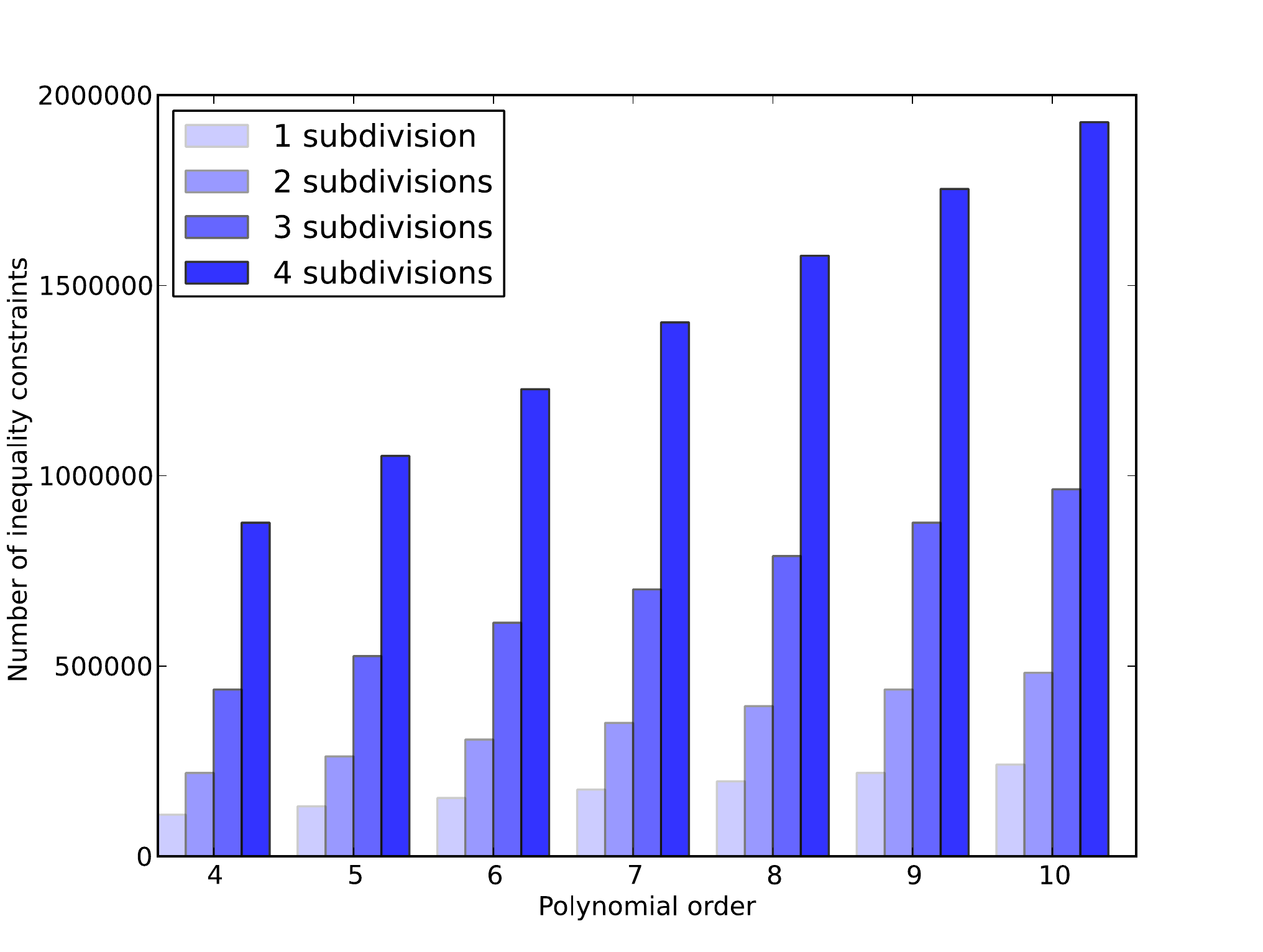}
\end{subfigure}
\begin{subfigure}[H]{0.45\textwidth}
\includegraphics[width = \textwidth]{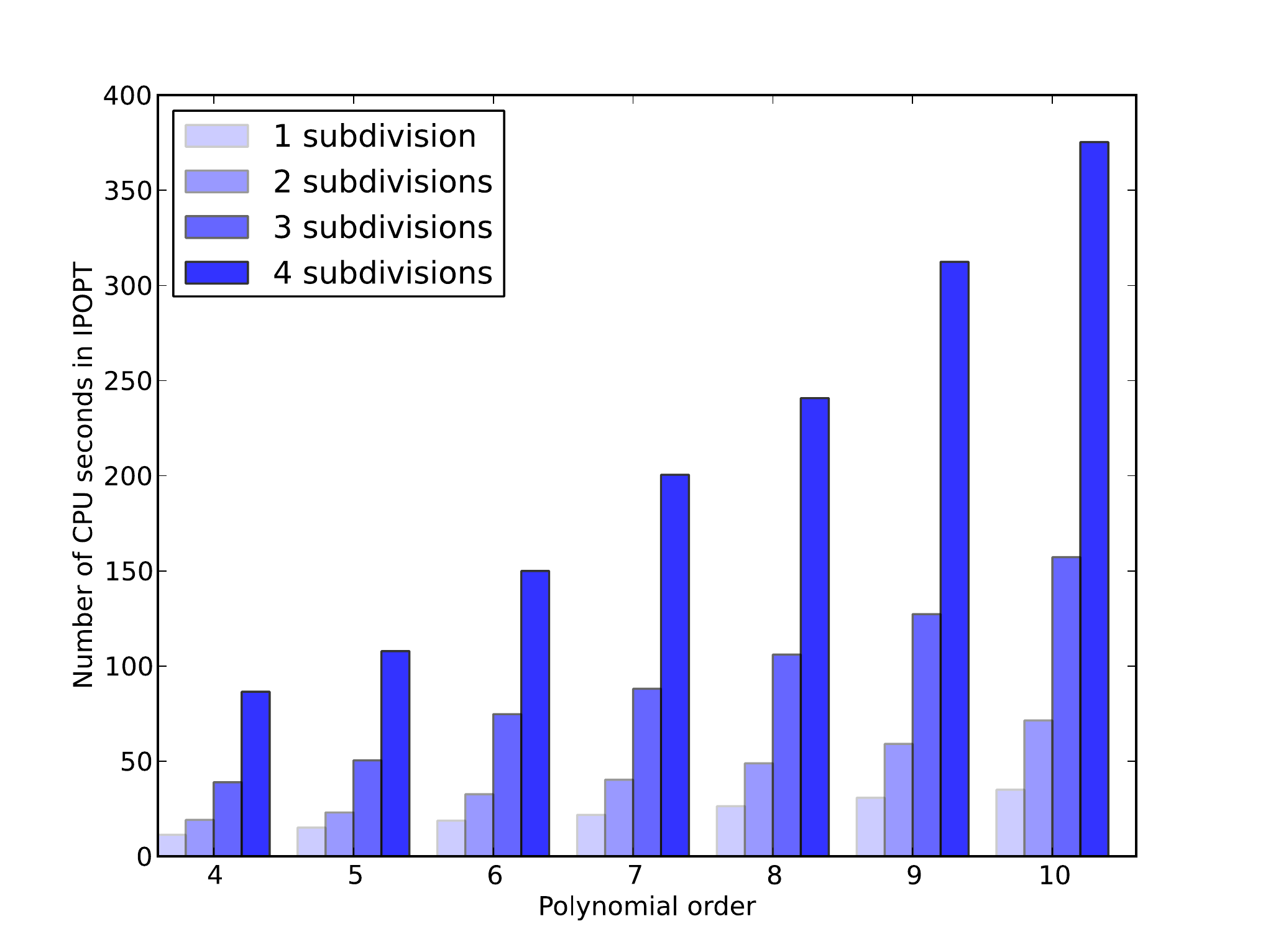}
\end{subfigure}
\caption{Bar plots comparing the objective function value (top left), the daily average Jensen-Shannon distance between original and synthesized data (top right), the number of inequalities in the problem formulation (bottom left) and the CPU time spent in IPOPT solving the optimization problem (bottom right) of the tested models.}
\label{fig3}
\end{figure}

Figure~\ref{fig3} shows bar plots comparing the different models using four criteria: the objective function value, the daily average Jensen-Shannon distance between original and synthesized wind direction data, the number of inequalities in the problem formulation and the CPU time spent in IPOPT solving the optimization problem. 
The Jensen-Shannon distance is the square root of the Jensen-Shannon divergence $d_{js}$, which, for two discrete probability distributions $q_1$ and $q_2$ is defined as:

\begin{equation*}
d_{js} = \frac{1}{2} \sum_i \frac{q_1(i)}{q_2(i)}\cdot q_1(i) + \frac{1}{2}\sum_i \frac{q_2(i)}{q_1(i)}\cdot q_1(i).
\end{equation*}

Comparing the models using the objective function value (figure~\ref{fig3} top left) shows a decrease of the objective function as the model order and the number of subdivisions increase. It can be seen that the impact of the number of subdivisions is higher for models with higher polynomial order. Moreover, the first subdivision has the highest impact since it leads to the highest decrease of the objective function value. The daily average of the Jensen-Shannon distance (figure~\ref{fig3} top right) decreases with the polynomial order until sixth order. The same behavior can be observed for the number of subdivisions: until the sixth order, the Jensen-Shannon distance decreases with the number of subdivisions. The number of inequality constraints in the optimization problem as well as the number of CPU seconds spent in the solver show the expected behavior (figure~\ref{fig3} bottom). They increase linearly with the polynomial order and exponentially with the number of subdivisions. 
Based on these observations, a basis order of 6 with 2 subdivisions was chosen as the best trade-off between an accurate representation of the average daily patterns and computational costs.

\subsection{Capturing long-term statistics}

This section compares the main statistical properties derived from the original data with the ones derived from the data generated by the time-variant Markov model.

\begin{figure}[H]
\begin{subfigure}[H]{0.32\textwidth}
\centering
\includegraphics[width = \textwidth]{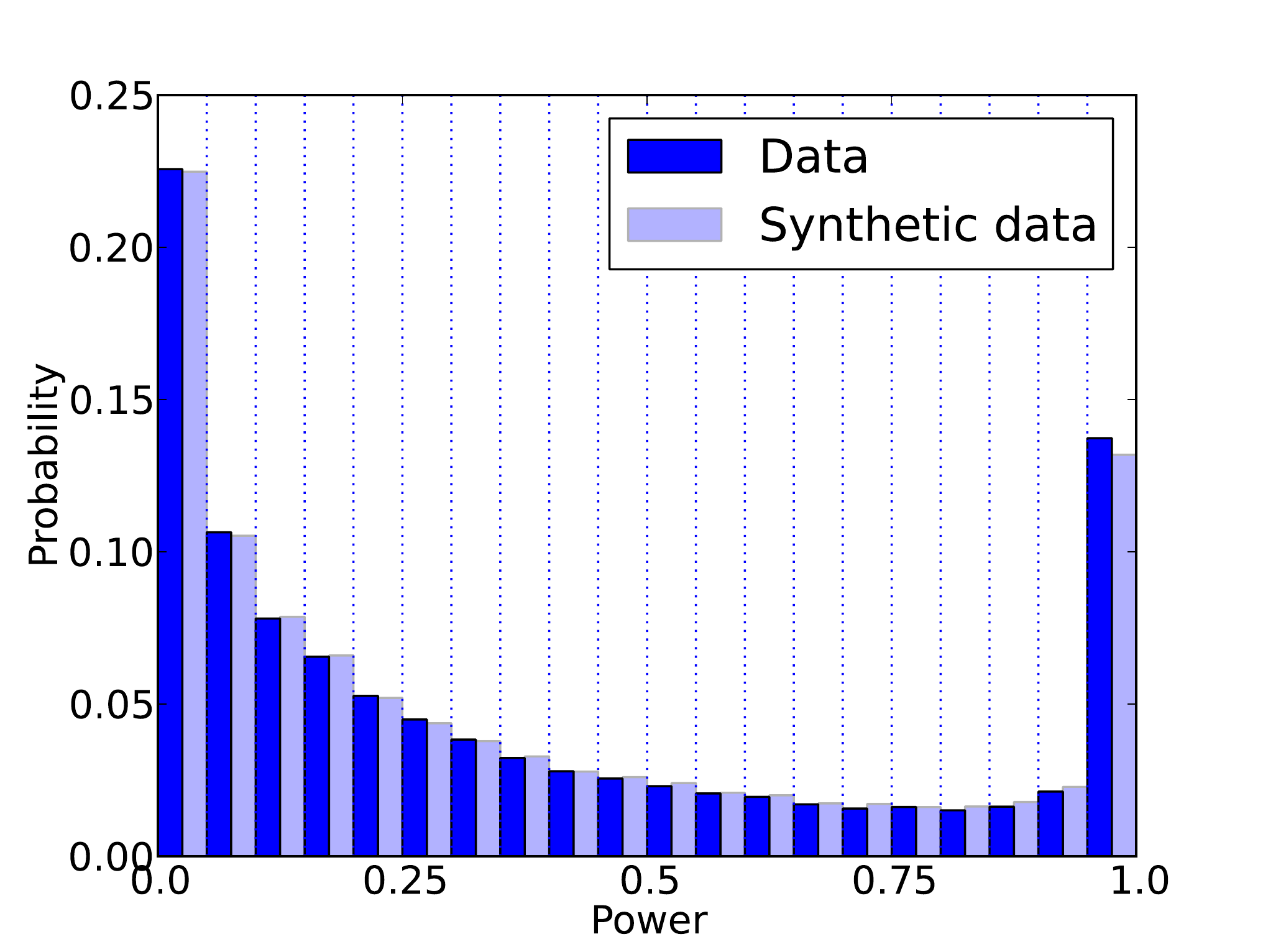}
\end{subfigure}
\begin{subfigure}[H]{0.32\textwidth}
\centering
\includegraphics[width = \textwidth]{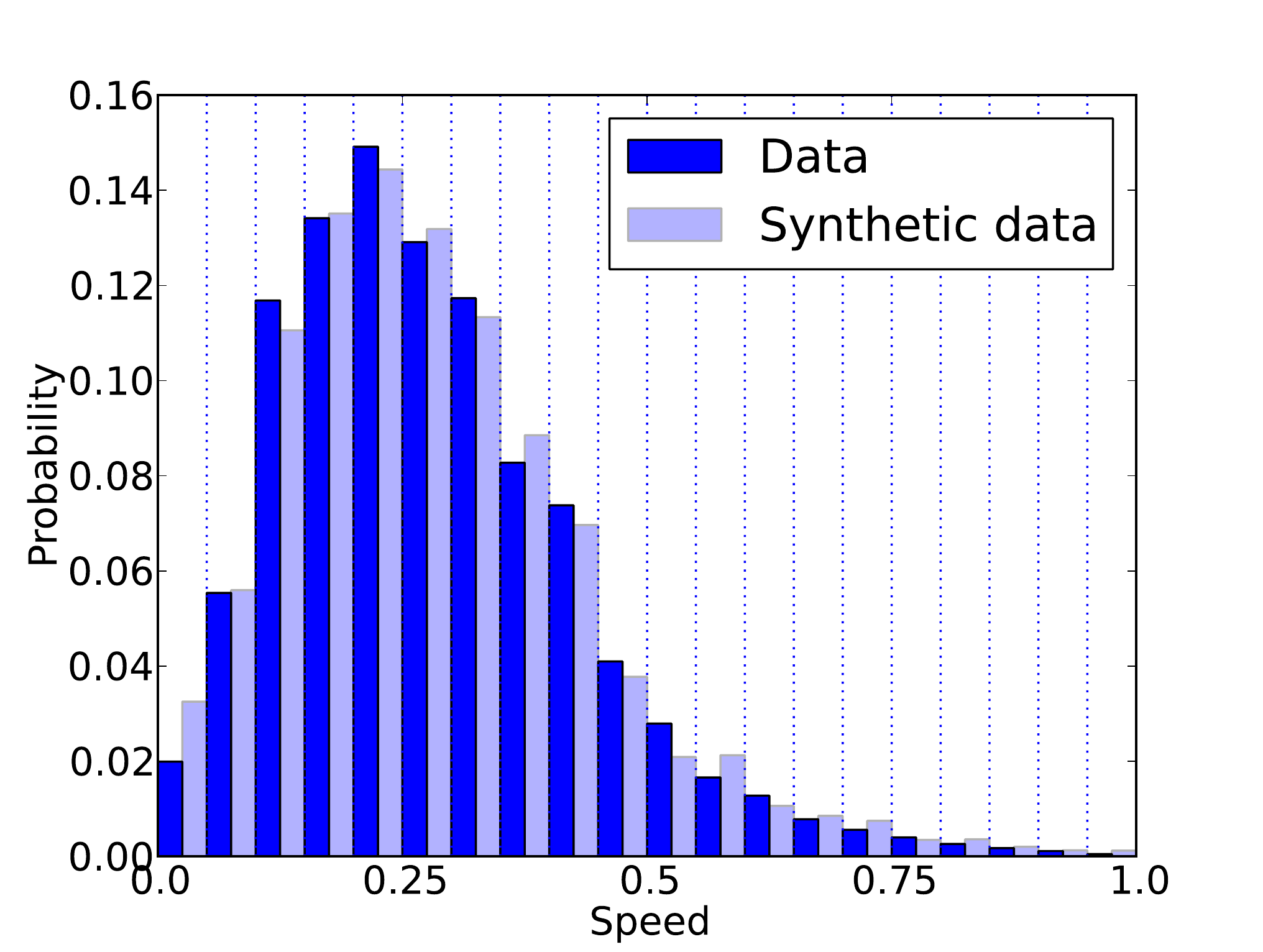}
\end{subfigure}
\begin{subfigure}[H]{0.32\textwidth}
\centering
\includegraphics[width = \textwidth]{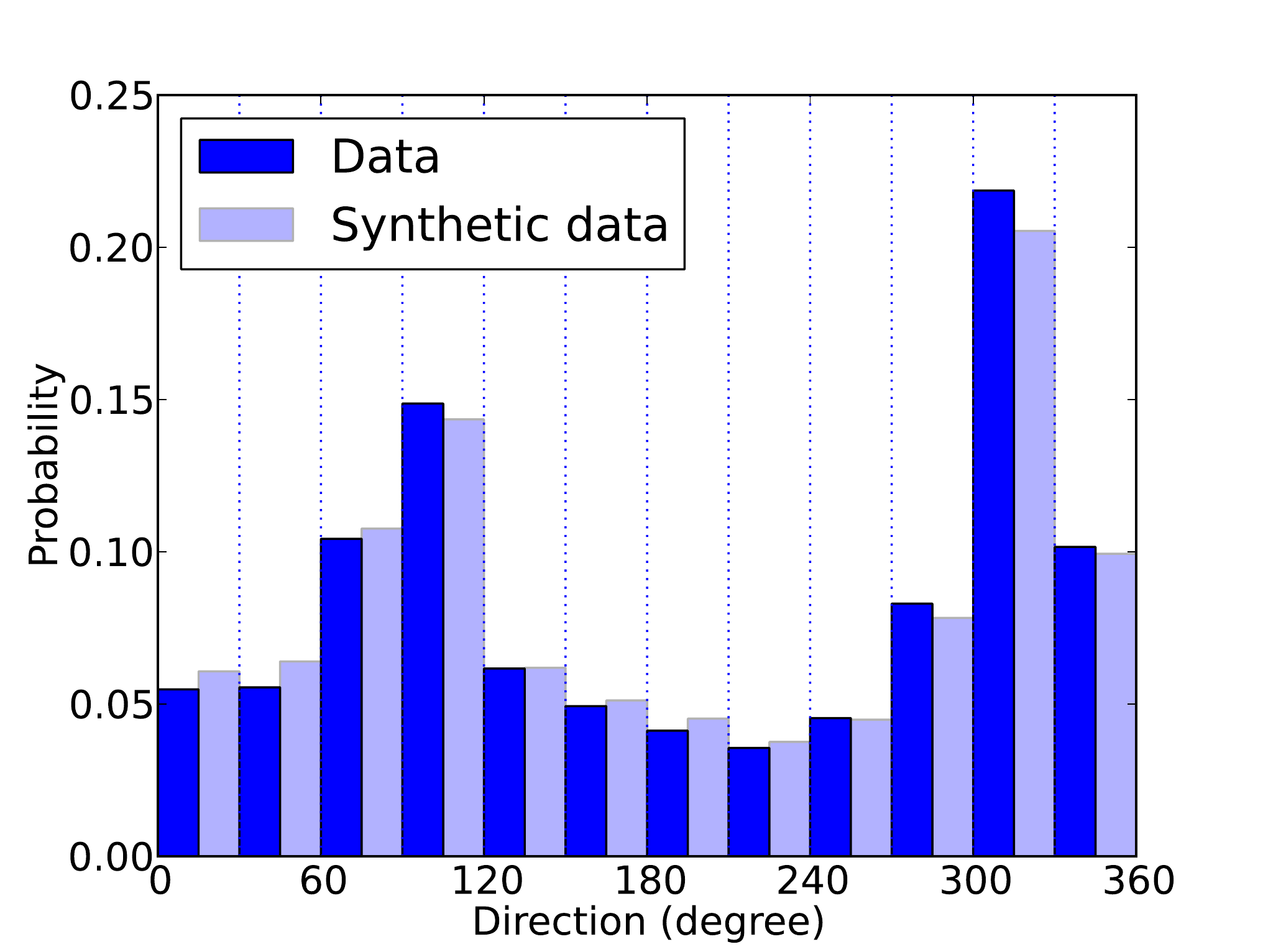}
\end{subfigure}
\caption{Comparison of the probability distribution of wind power (left), wind speed (middle) and wind direction (right) of the original with the synthesized data.}
\label{fig4}
\end{figure}

Figure~\ref{fig4} compares the wind power (left), speed (middle) and direction (right) distribution of the original with the synthetic data generated using the Markov model. In general, the distributions are in close agreement. The wind power distribution is bimodal, with the modes located at the minimum and maximum power. It shows that the intermediary power levels are rather rare, for instance, the states corresponding to a power production between 0.4 and 0.9 have a low probability. The wind speed distribution follows the expected behavior, a single mode distribution with a long tail for the high wind speeds (Weibull distribution). The wind direction distribution is bimodal with the two modes at 100 and 300 degrees, which are the prevailing wind directions at the turbine site (figure~\ref{fig2}).\\

\begin{figure}[H]
\begin{subfigure}[H]{0.49\textwidth}
\centering
\includegraphics[width = \textwidth]{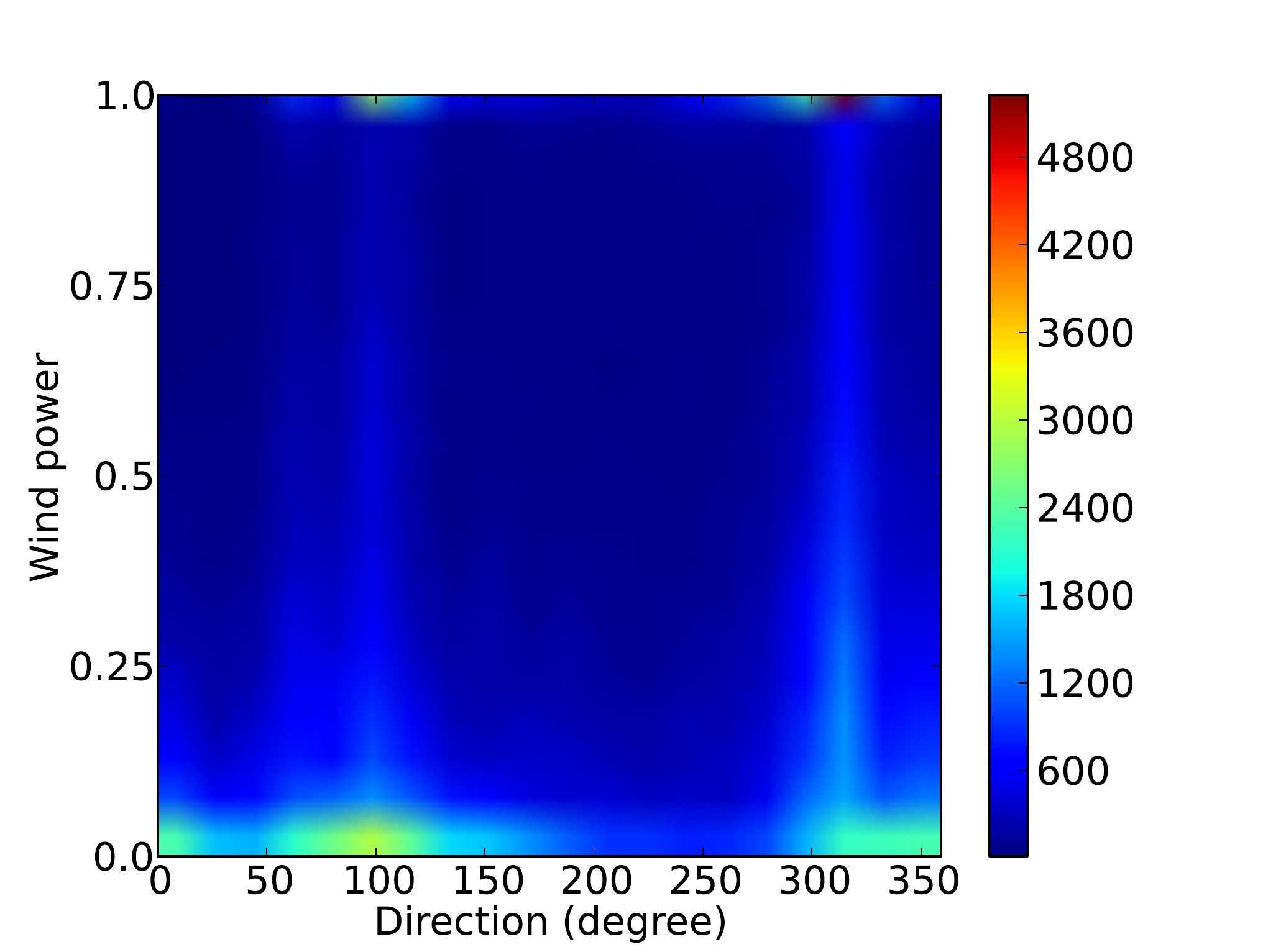}
\end{subfigure}
\begin{subfigure}[H]{0.49\textwidth}
\centering
\includegraphics[width = \textwidth]{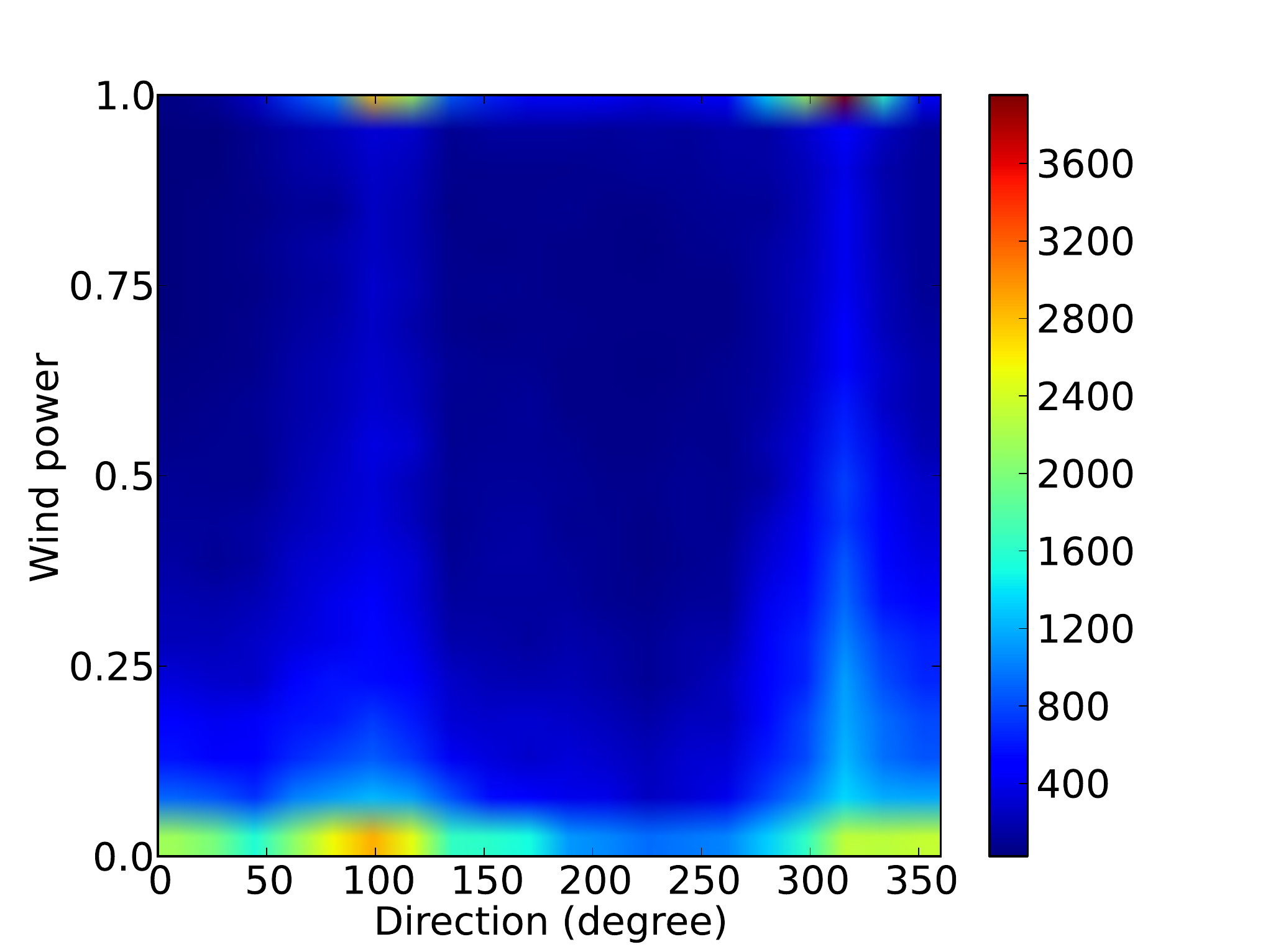}
\end{subfigure}
\caption{Two dimensional power-time histograms of the original (left) and synthetic (right) time-series data.}
\label{fig5}
\end{figure}

Figure~\ref{fig5} shows two plots: on the left, the empirical 2D distribution of the wind power and direction computed from the data and, on the right, the same distribution computed using the data generated by the Markov model. Its comparison shows that the model captures the joint statistics for the wind power and direction from the data. It is possible to see the two dominant directions associated with high wind power production, namely the sectors from 100 to 120 and from 290 to 320 degree. Figures~\ref{fig2} and~\ref{fig5} clearly demonstrate the capability of this Markov model to capture the combined characteristics of the wind power, speed and direction. The long-term behavior of the model is close to what is observed in the dataset.

\subsection{Capturing time-dependent behavior}

As shown in section~\ref{dailypatterns}, the original data clearly exhibits a time-dependent behavior. To test, if the time-dependent Markov model can capture it, synthetic data was generated and the histograms compared to the ones of the original data. Moreover, to obtain a comparison with the ``regular'' way of data synthesis with Markov models, data was also generated from the time-invariant Markov chain.

\begin{figure}[H]
\centering
\begin{subfigure}[H]{0.45\textwidth}
\includegraphics[width = \textwidth]{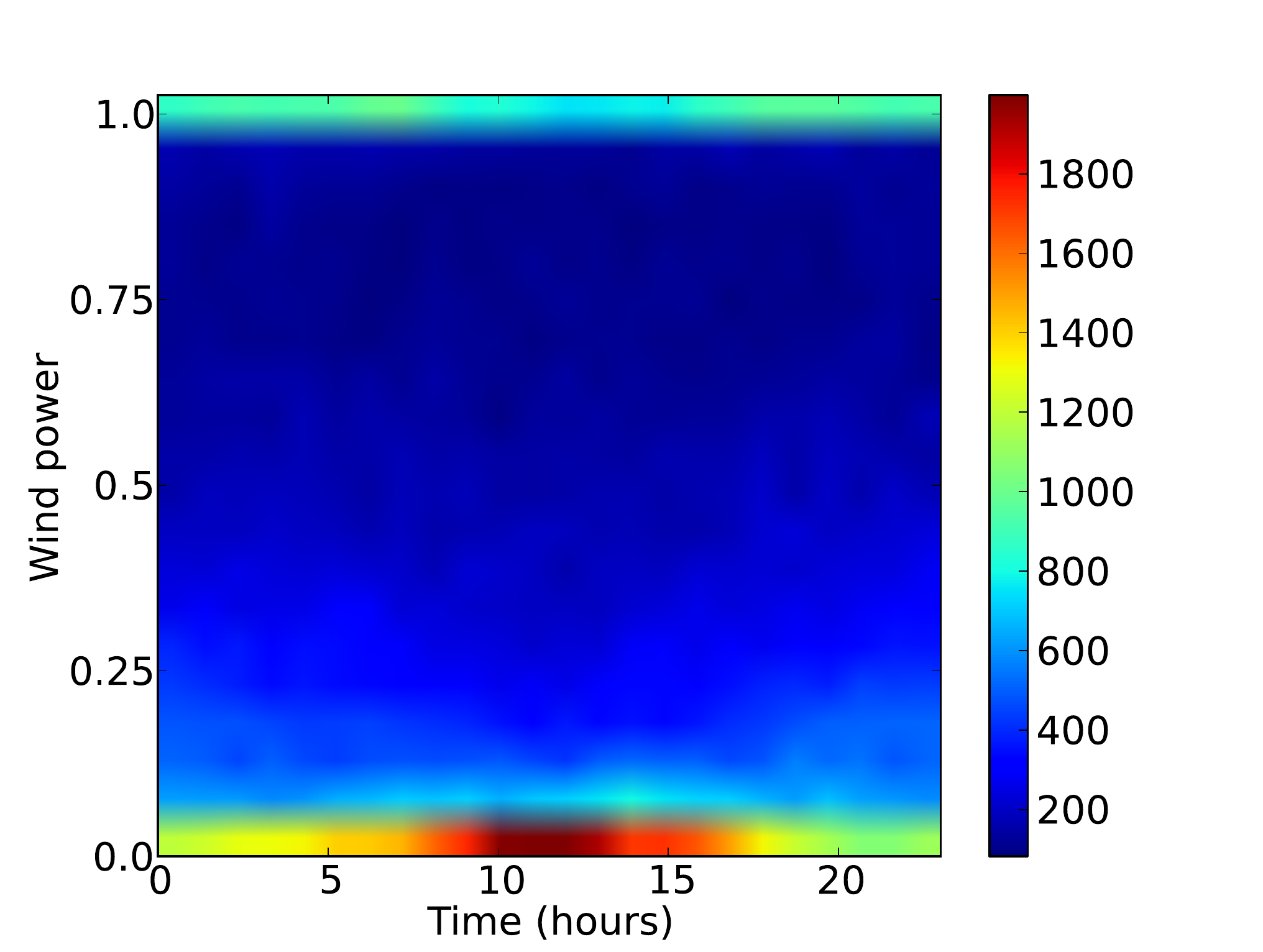}
\end{subfigure}
\begin{subfigure}[H]{0.45\textwidth}
\includegraphics[width = \textwidth]{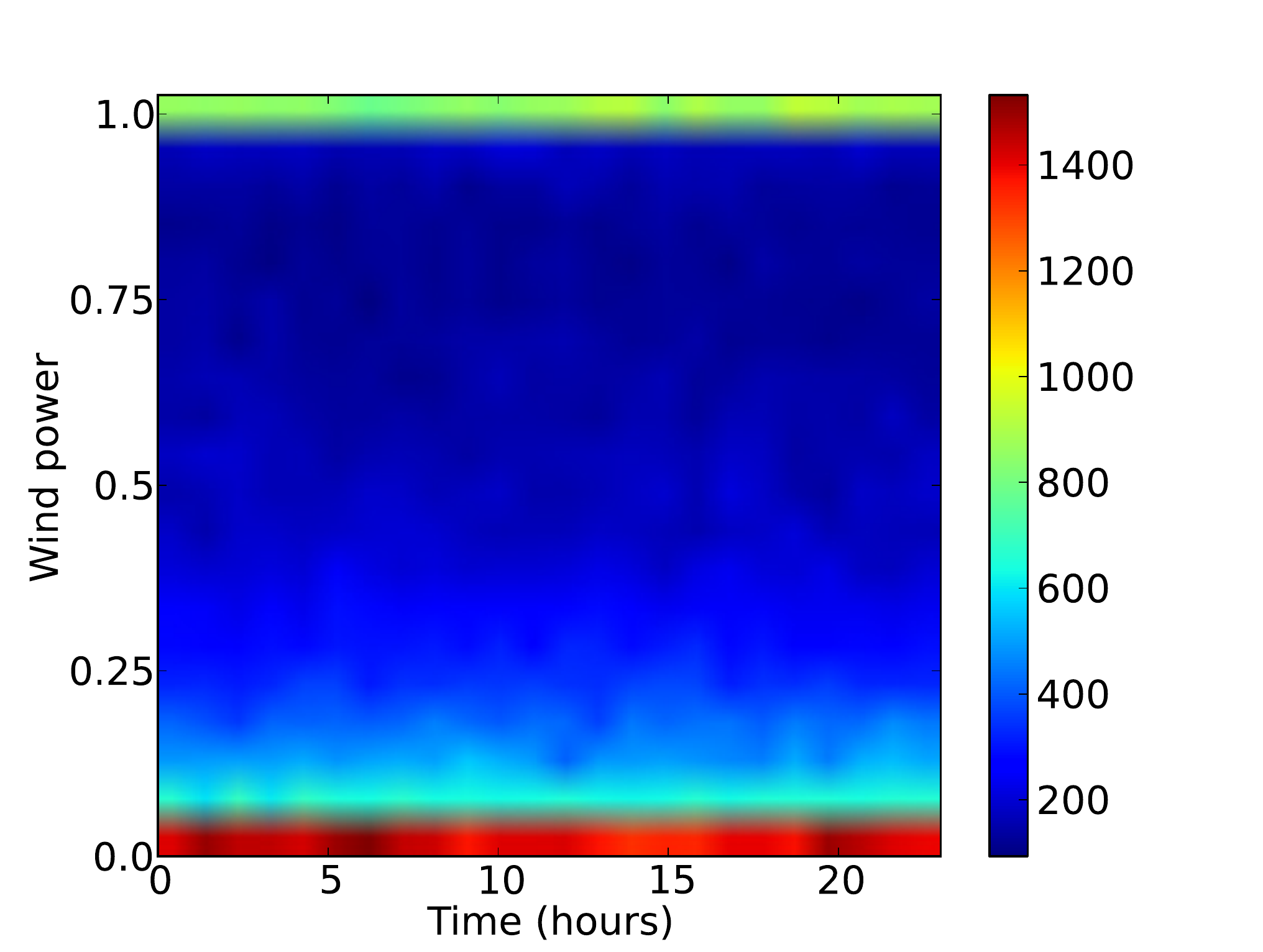}
\end{subfigure}\\
\begin{subfigure}[H]{0.45\textwidth}
\includegraphics[width = \textwidth]{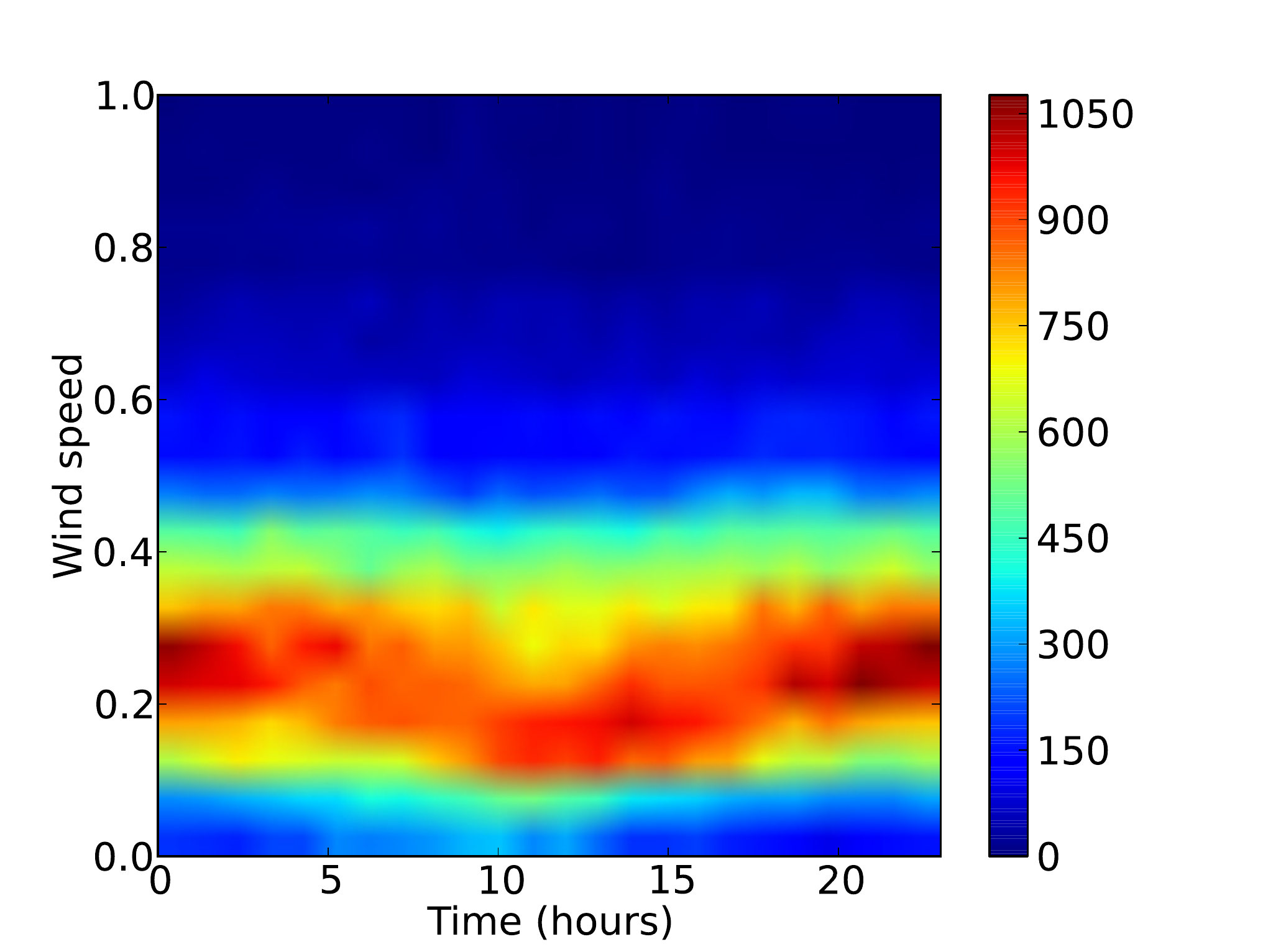}
\end{subfigure}
\begin{subfigure}[H]{0.45\textwidth}
\includegraphics[width = \textwidth]{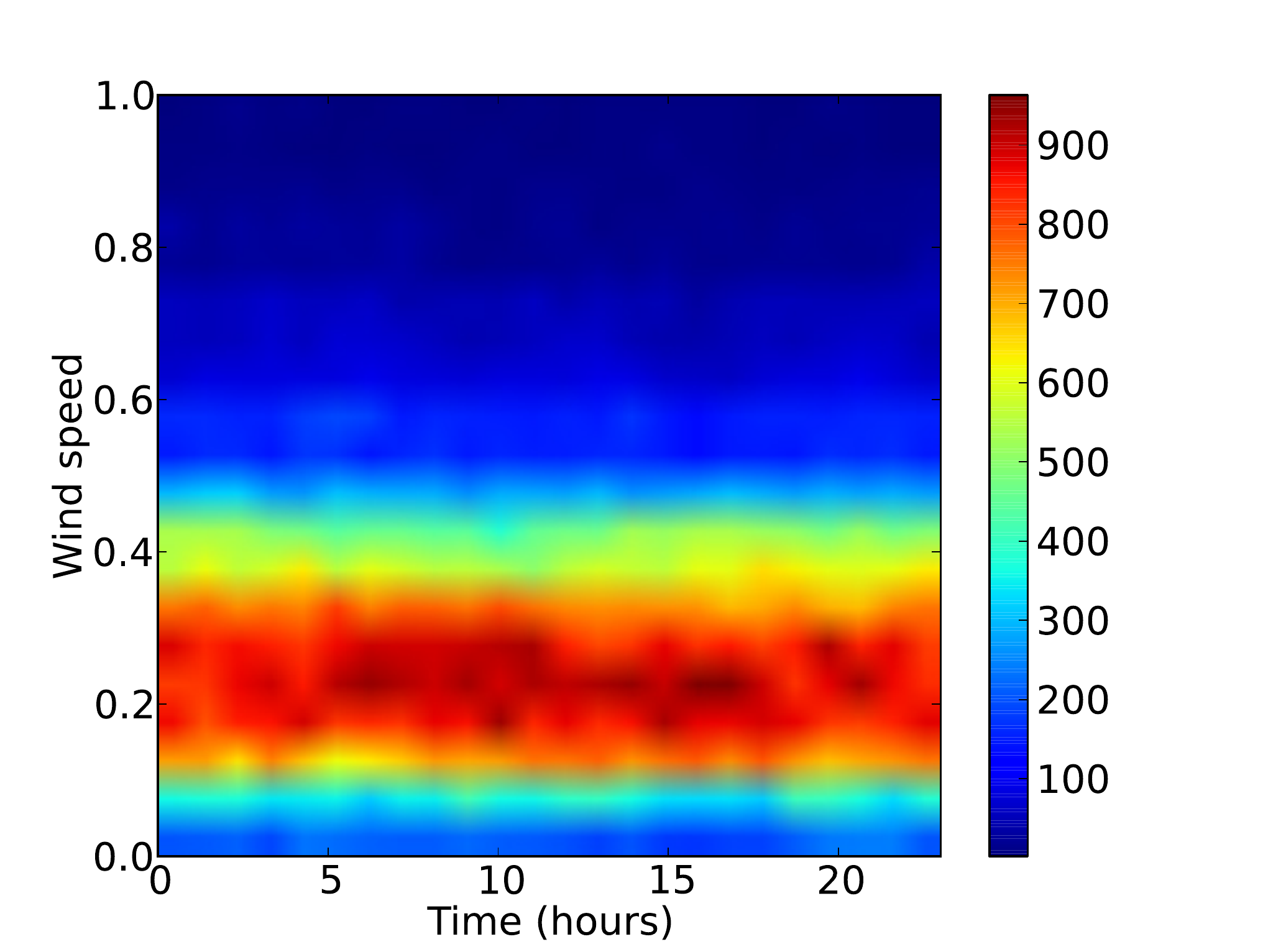}
\end{subfigure}\\
\begin{subfigure}[H]{0.45\textwidth}
\includegraphics[width = \textwidth]{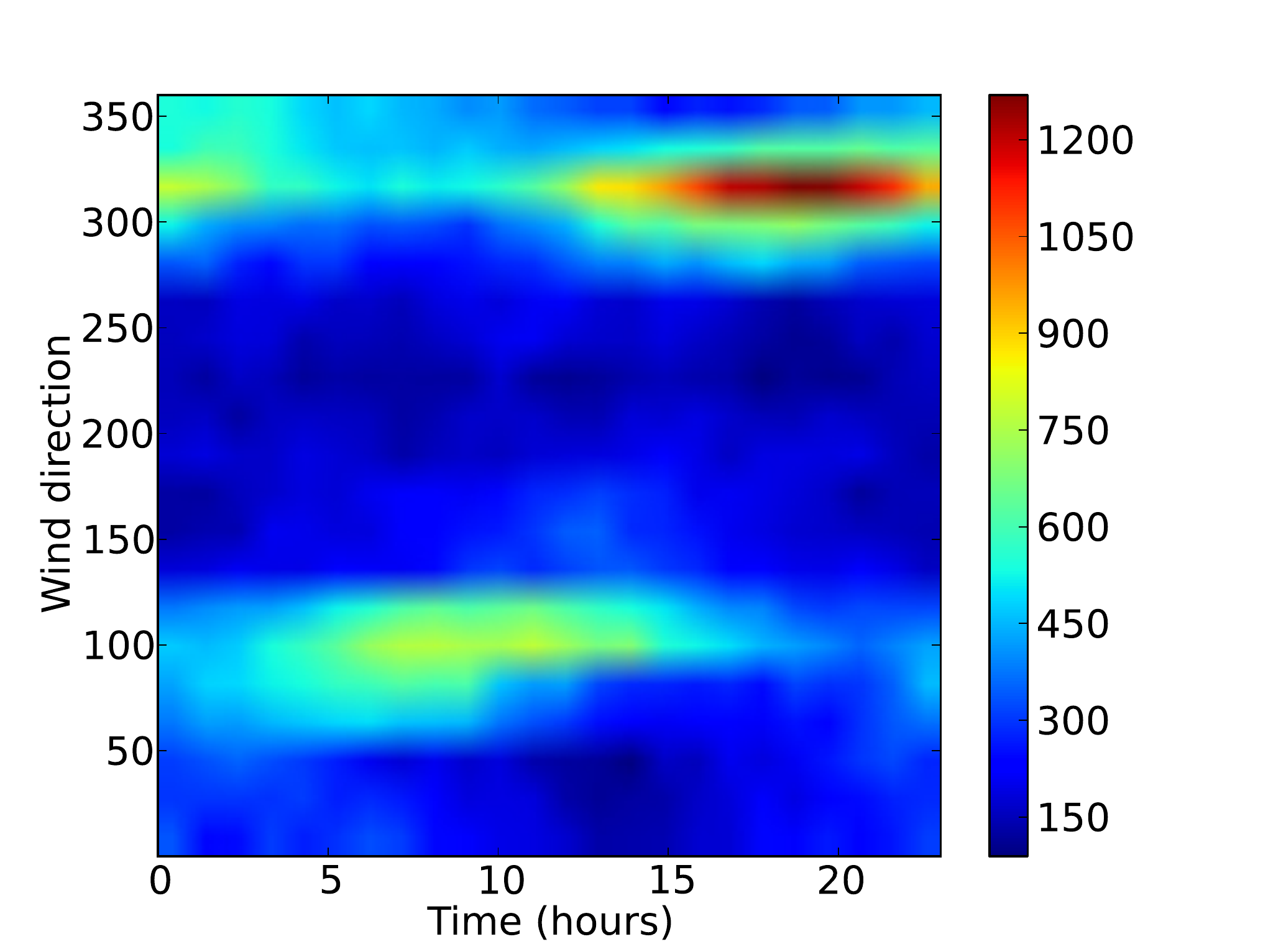}
\end{subfigure}
\begin{subfigure}[H]{0.45\textwidth}
\includegraphics[width = \textwidth]{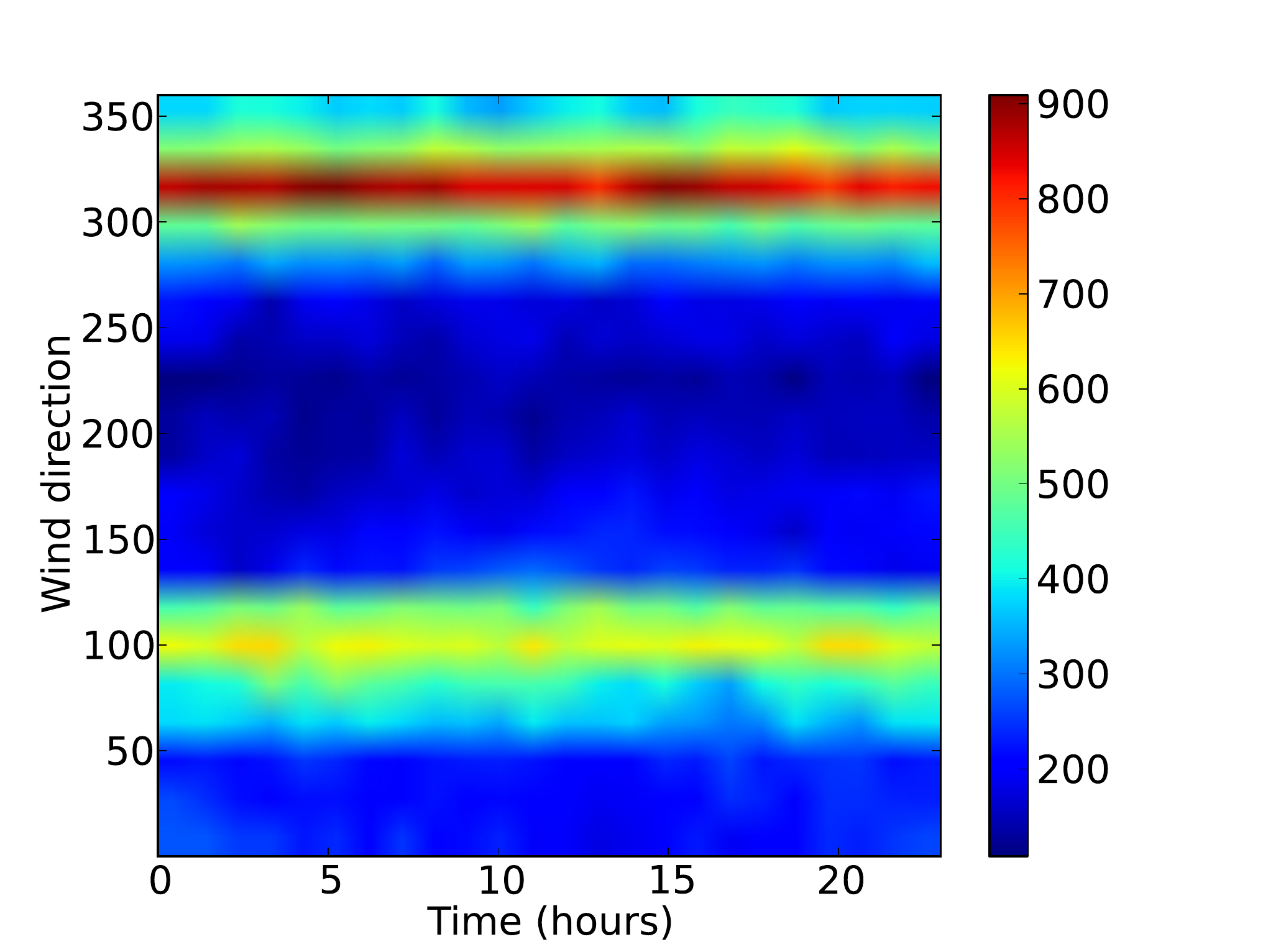}
\end{subfigure}
\caption{Two dimensional histograms of the synthetic time-series data, generated with the time-variant Markov model (left) and the time-invariant Markov chain (right): power-time (top), speed-time (middle) and direction-time (bottom).}
\label{fig6}
\end{figure}

The comparison of figures~\ref{fig2} and~\ref{fig6}(first column) shows, that the time-variant Markov model is capable of reproducing the time-variant behavior of the data. Figure~\ref{fig6}(second column) presents the results of using a time-invariant Markov chain model, i.e. by using constant transition probability functions. As expected, each variable statistic distribution remains constant during the daily cycle.

\subsection{Time-dependent persistence of production}

The time-dependent Markov model allows to compute the persistence of power-production depending on the time of the day. Figure~\ref{fig6} shows the time-dependent persistence of power production for different power levels ($i \cdot 0.05 \cdot p_{max}$, for $i = 1,...,19$). The persistence analysis is presented for two power production levels: a) persistence of useful power production (PUPP) defined as above $0.15 \cdot p_{max}$, i.e. the power level corresponding to the wind speed mode at the turbine site; and, persistence for high power production (PHPP), i.e. above $>0.7 \cdot p_{max}$. It can be seen, that the higher the power level, the lower the persistence. Moreover, for all power levels, persistence is minimal between 5 and 10 am. PHPP is fairly constant throughout the day (dark line), the maximal differences are between 10 and 30 minutes, whereas PUPP reaches a maximum at around 5 pm (white line).

\begin{figure}[H]
\centering
\includegraphics[width = \textwidth]{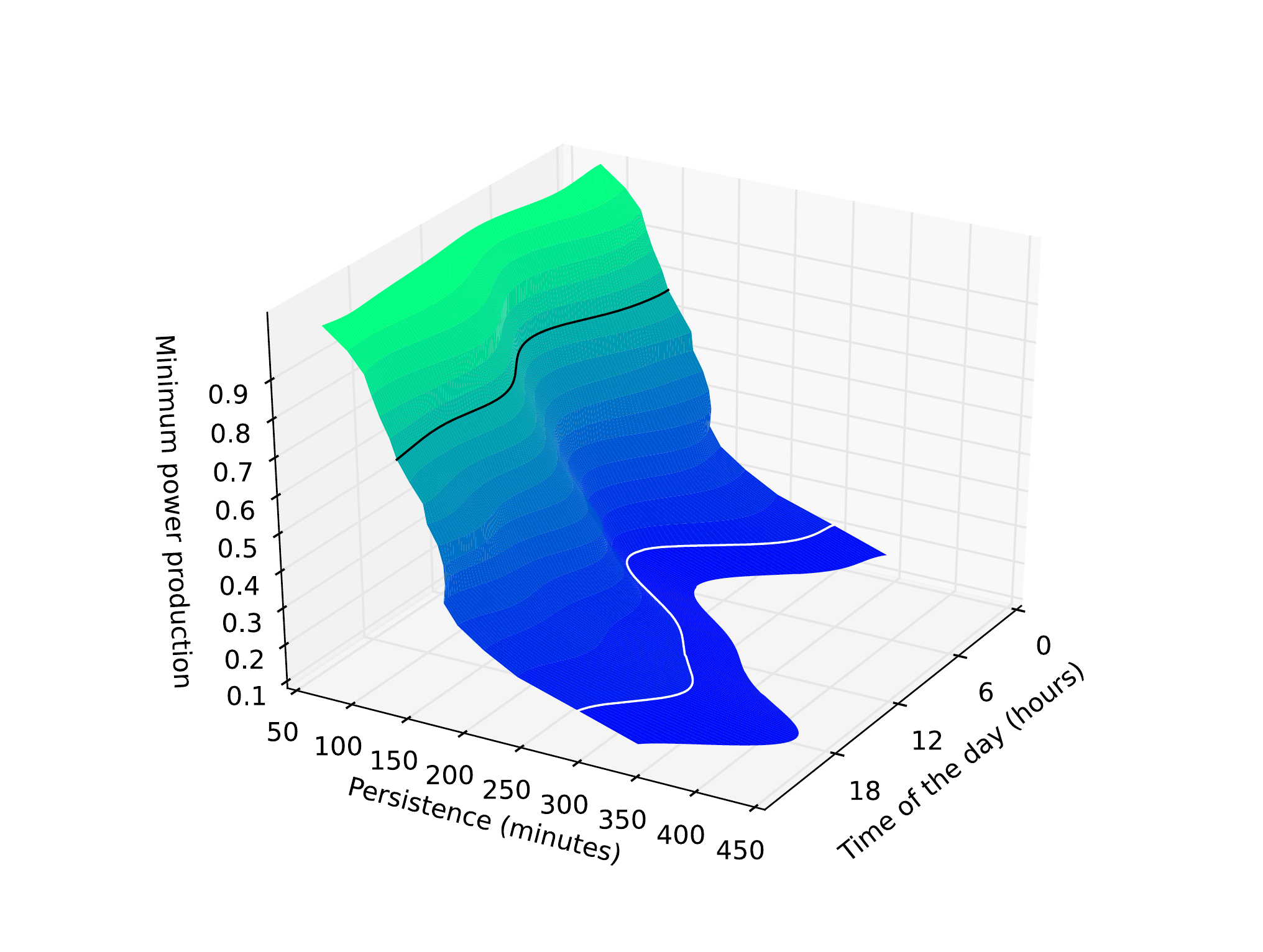}
\caption{Time-dependent persistence of power production above $i \cdot 0.05 \cdot p_{max}$, for $i = 1,...,19$. The lines highlight the time-dependent persistence, for two conditions: a) in white, for power production above $0.15 \cdot p_{max}$ (PUPP); and, b) in black, the power production above $0.7 \cdot p_{max}$ (PHPP).}
\label{fig7}
\end{figure}

\begin{figure}[H]
\begin{subfigure}[H]{0.48\textwidth}
\centering
\includegraphics[width = \textwidth]{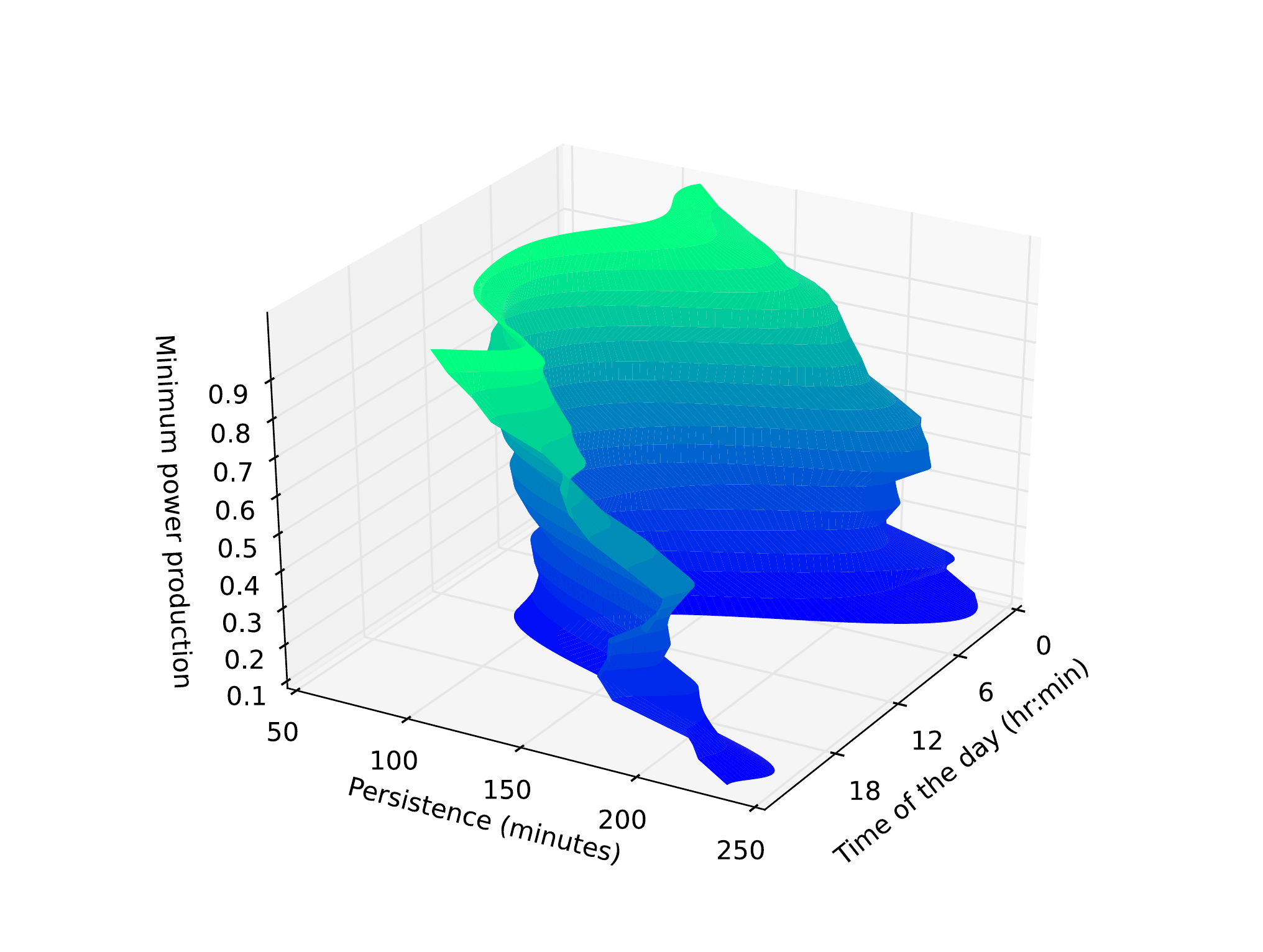}
\end{subfigure}
\begin{subfigure}[H]{0.48\textwidth}
\centering
\includegraphics[width = \textwidth]{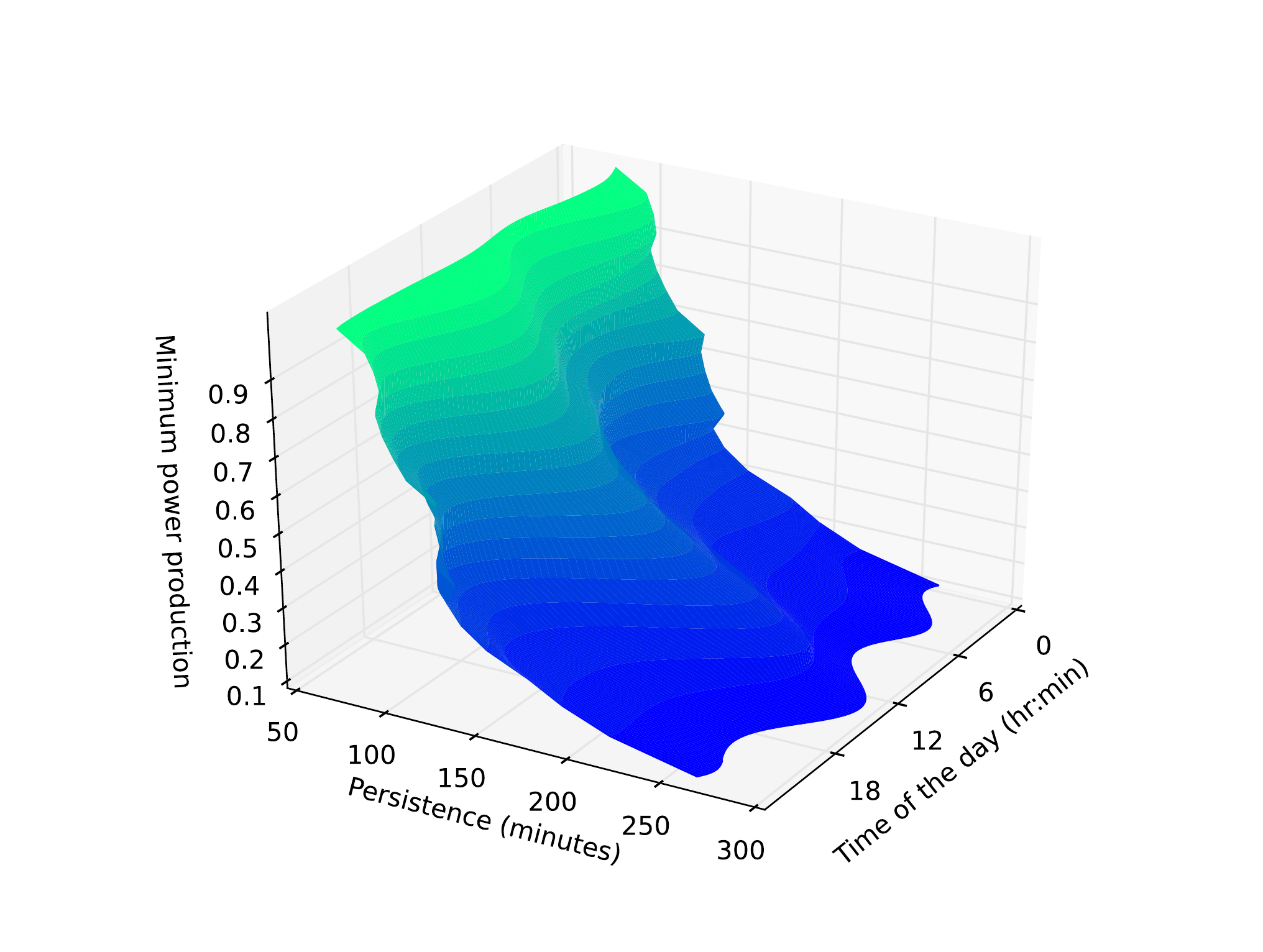}
\end{subfigure}
\caption{Time-dependent persistence of power production above $i \cdot 0.05 \cdot p_{max}$, for $i = 1,...,19$ for direction sectors $90^{\circ}$-$180^{\circ}$ (left) and $270^{\circ}$-$360^{\circ}$ (right).}
\label{fig8}
\end{figure}

Since the data shows two different dominant directions (figure~\ref{fig4}), figure~\ref{fig8} presents the persistence of power production conditioned to each dominant direction, i.e. for the direction sectors from $90^{\circ}$ to $180^{\circ}$ and  $270^{\circ}$ to $360^{\circ}$.

As expected, the persistence in both direction sectors is lower than the unconditioned persistence. For wind directions in the sector $90^{\circ}$-$180^{\circ}$, all levels of power production have a minimum persistence between 80 and 100 minutes at around 1 pm. Maximum persistence is around midnight varying between 220 minutes (PUPP) and 140 minutes (PHPP). It can be seen that for power levels below 50\% of maximum production the time of day dependency of persistence is very similar. For power levels above 50\% persistence decreases as power production increases. However, the persistence variability with the power level is rather low, for example, maximum persistence at a level of 75\% is almost 180 minutes whereas for a level above 0.05\% is 200 minutes.\\
For wind directions in the sector $270^{\circ}$-$360^{\circ}$, it shows that, for all power levels, the curves for both PUPP and PHPP are similar, i.e. their minima and maxima are located around the same time of the day. For instance, maximal persistence of production is reached at around midday. However, for this direction sector, the higher the power production level, the lower the persistence. For power production above 0.05\% of maximum power the persistence is 250 minutes, persistence of production above 75\% of maximum power is only 100 minutes.\\
Comparing with the other dominant direction, it can be seen, that they have very different persistence behavior. The maxima and minima are at different times of the day for every power level. The persistence increases as power production decreases, for all power levels in the case of the $270^{\circ}$-$360^{\circ}$ sector. For the $90^{\circ}$-$180^{\circ}$ sector it decreases only until 50\% of maximum power production. Below that, it remains approximately constant.

\section{Conclusions}
\label{Conclusions}

This paper presents an inhomogeneous Markov process to model wind power production. It is developed using states, which combine information about the wind speed, direction and power variables, using real data recorded by a wind turbine in Portugal. The joint partition of the three-dimensional variable space allows to decrease the number of the model states and, simultaneously, encodes the wind power curve into the Markov chain model. The transition probabilities are considered to be functions that depend on the time of the day and modeled as Bernstein polynomials. The estimation of the transition matrices is performed by solving a constrained convex optimization problem. Its objective function combines two log-likelihood functions with the purpose to accurately represent both the long-term behavior and the daily fluctuations seen in the original data.
To evaluate the statistical properties of the estimated Markov model, synthetic time-series are generated and compared with the original data statistics. Results demonstrate that the proposed Markov model can reproduce the diurnal patterns in the data. Moreover it is demonstrated how the persistence of power production throughout the time of the day can be estimated from the Markov process transition matrices.

\section*{Acknowledgments}

The authors thank the Funda\c{c}\~ao para a Ci\^encia e a Tecnologia for financial support (SFRH/BD/86934/2012, FCOMP-01-0124-FEDER-016080 (PTDC/SENENR/1141718/2009)) and GENERG, SA.





\newpage
\bibliographystyle{elsarticle-harv}
\bibliography{windbib.bib}







\end{document}